\documentclass[twocolumn]{aastex63}

\accepted{for publication in ApJ, July 2021}

\shorttitle{X-ray Super-Flares}
\shortauthors{Getman, Feigelson, \& Garmire}

\graphicspath{{./}}

\begin{document}

\title{X-ray Super-Flares From Pre-Main Sequence Stars: Flare Modeling}

\correspondingauthor{Konstantin Getman}
\email{kug1@psu.edu}

\author[0000-0002-6137-8280]{Konstantin V. Getman}
\affiliation{Department of Astronomy \& Astrophysics \\
Pennsylvania State University \\ 
525 Davey Laboratory \\
University Park, PA 16802, USA}

\author{Eric D. Feigelson}
\affiliation{Department of Astronomy \& Astrophysics \\
Pennsylvania State University \\ 
525 Davey Laboratory \\
University Park, PA 16802, USA}

\author{Gordon P. Garmire}
\affiliation{Huntingdon Institute for X-ray Astronomy\\
LLC, 10677 Franks Road\\
Huntingdon, PA 16652, USA}

\begin{abstract}
Getman et al. (2021) reports the discovery, energetics, frequencies, and effects on environs of $>1000$ X-ray super-flares with X-ray energies $E_X \sim 10^{34}-10^{38}$~erg from pre-main sequence (PMS) stars identified in the  $Chandra$ MYStIX and SFiNCs surveys. Here we perform detailed plasma evolution modeling of $55$ bright MYStIX/SFiNCs super-flares from these events. They constitute a large sample of the most powerful stellar flares analyzed in a uniform fashion. They are compared with published X-ray super-flares from young stars in the Orion Nebula Cluster, older active stars, and the Sun. Several results emerge.  First, the properties of PMS X-ray super-flares are independent of the presence or absence of protoplanetary disks inferred from infrared photometry, supporting the solar-type model of PMS flaring magnetic loops with both footpoints anchored in the stellar surface. Second, most PMS super-flares resemble solar long duration events (LDEs) that are associated with coronal mass ejections. Slow rise PMS super-flares are an interesting exception. Third, strong correlations of super-flare peak emission measure and plasma temperature with the stellar mass are similar to established correlations for the PMS X-ray emission composed of numerous smaller flares. Fourth, a new correlation of loop geometry is linked to stellar mass; more massive stars appear to have thicker flaring loops.  Finally, the slope of a long-standing relationship between the X-ray luminosity and magnetic flux of various solar-stellar magnetic elements appears steeper in PMS super-flares than for solar events. 
\end{abstract}

\section{Introduction} \label{sec:intro2}

The contemporary Sun exhibits mild magnetic activity with flares of low and moderate power.  The physics of solar flares is reasonably well-understood, aided by high resolution movies of the events in different wavebands including radio, H$\alpha$, ultraviolet, X-rays, and magnetograms \citep[see reviews by][]{Shibata2011, Benz2017}.  For the most powerful solar  flares, explosive magnetic reconnection occurs at the top of cusp-shaped magnetic loops, releasing accelerated particles that precipitate to the chromosphere, evaporating plasma that fills the loop with X-ray emitting plasma.  The X-ray luminous class of long-duration events (LDEs) have characteristic durations ranging from an hour to a day, associated with coronal loops reaching heights up to $10\%-50$\% of the Sun's radius with total radiated energies ranging from $\log E_X \sim 30 - 32$ ergs. 

For a wide range of solar flare energies, a distinctive multi-stage behavior of the X-ray emitting flare plasma is often seen: Phase I with the sudden injection of heat, Phase II with the rise to a maximum temperature as the loop fills with evaporated material, Phases III and IV when conductive and radiative processes cool the plasma \citep{Reale2014}. 

The most powerful stellar flares are found in stars during their pre-main sequence (PMS) evolutionary phase.  With rapid rotation and fully convective interiors, magnetic dynamos efficiently generate fields \citep{Yadav2015, Warnecke2020} that cover a large fraction of the stellar surface with active regions \citep{GullySantiago2017, Kochukhov2020,Morris2020}.  The most powerful PMS flares are nicknamed `super-flares’ or `mega-flares', with peak luminosities in the range $\log L_X \sim 31 - 33$ erg~s$^{-1}$ and total radiated energies $\log E_X \sim 34 - 38$ erg in the X-ray band.  

It has not been clear whether the solar analogy should be applied at the early stages of PMS evolution because the magnetosphere is believed interact with the protoplanetary disk.  Magnetic fields truncate the inner disk, funneling some disk material to the stellar surface and ejecting other material in a wind or jet \citep{Gregory2010, Johnstone2014, Romanova2015}.  Associated transfer of angular momentum slows the stellar rotation.  Stellar field lines that penetrate into, or are confined by, a differentially rotating disk are predicted to  undergo violent recombination and flaring close to the disk \citep{Hayashi1996, Shu1997, Takasao2019, Colombo2019}.  Careful comparison of X-ray flaring in disk-bearing and diskless populations is needed to test these models.  

In our previous study \citep[][hereafter Paper I]{Getman2021}, we extract 1,086 super- and mega-flares from X-ray lightcurves of $\sim$24,000 stars with ages $\leq 5$~Myr associated with several dozen Galactic molecular clouds observed with NASA’s {\it Chandra X-ray Observatory}. These stars are members of the MYStIX \citep[Massive Young Star-forming complex study in Infrared and X-rays;][]{Feigelson13} and SFiNCs \citep[Star Formation in Nearby Clouds;][]{Getman18b} archive studies of star forming regions.  They are associated with underlying populations totaling $\sim 112,000$ pre-main sequence (PMS) stars down to $0.1$~M$_{\odot}$.  PMS flaring is readily studied with {\it Chandra} as hundreds or thousands of young stellar objects can be observed simultaneously in a {\it Chandra} exposure of a rich young star cluster. In Paper I, we call events with X-ray energies $10^{34} < E_X < 10^{36.2}$~erg `super-flares' and events with $E_X > 10^{36.2}$~erg `mega-flares'. For simplicity, we will hereafter call MYStIX/SFiNCs flares as `super-flares', bearing in mind that the majority of the  MYStIX/SFiNCs super-flares modeled here are actually mega-flares.

The MYStIX/SFiNCs super- and mega-flares are not the first stellar flares known with this power. \citet{Gudel2004} lists $\sim 30$ stellar flares of comparable power observed with different instruments prior to {\it Chandra}. \citet{Colombo07} identifies over 900 X-ray flares from the 13-day {\it Chandra} observation of the Orion Nebula Cluster, the {\it Chandra} Orion Ultradeep Project \citep[COUP;][]{Getman05}. Of those, the brightest 32 and 216 COUP super-flares were modeled by \citet{Favata2005} and \citet{Getman08a}, respectively.

Other {\it Chandra} and {\it XMM-Newton} studies of less populous flare samples from young stellar objects include \citet{Grosso2004, Favata2005, Wolk05, Argiroffi2006, Giardino2006, Colombo07, Caramazza07, Giardino2007, Franciosini2007, Stelzer07, LopezSantiago2010, McCleary2011,Bhatt2014, Flaccomio2018, Guarcello2019, Pillitteri2019} and \citet{Grosso20}.  The targets of these studies are star forming regions generally closer than 0.5~kpc rather than the more distant regions in the MYStIX/SFiNCs surveys.  

The MYStIX/SFiNCs surveys have large \'etendue and thus provide the largest available sample of super-flares that we identify and analyze in a uniform fashion. Paper~I shows that these X-ray super-flares are produced by young stars spanning a wide mass range from M- to OB-type stars.  The distribution of super-flare energies follows a $dN/dE_{X} \propto E_X^{-2}$ relation regardless the presence/absence of circumstellar disks.  This power law distribution is similar to those of flares from older stars and the Sun. Mega-flares with $\log(E_X) > 36.2$~erg from solar-mass stars have occurrence rate of $1.7_{-0.6}^{+1.0}$ flares/star/year and contribute at least $10-20$\% to the total PMS X-ray energetics. The super-flare rate scales with stellar mass. Such powerful flares will have a significant effect on gas removal from the protoplanetary disks by photo evaporation. They may affect planet formation processes and primordial planetary atmospheres.

In the present study, we examine a subset of 55 MYStIX/SFiNCs super-flares that typically dominate their $>1000$-photon host {\it Chandra} observations (Appendix \ref{sec:photon_arrival}) and are sufficiently bright for analyses of flare temporal history. The phases of heating and cooling described by \citet{Reale2014} can be traced in these X-ray lightcurves allowing estimation of loop heights, geometry and cooling processes using methods developed in \citet{Getman2011}. We combine the 55 MYStIX/SFiNCs super-flares with  216 COUP super-flares from \citet{Getman08a}. These 271 MYStIX/SFiNCs/COUP super-flares represent the most energetic flares from PMS stars ever modelled, so our findings give an objective portrait of the super-flare phenomenon.  The host stars span a wide range of masses, both with and without protoplanetary disks.  We can thus test the solar flare paradigm against star-disk interaction models of flare generation.   

We find here that the evolution of most PMS super-flares is similar to that seen in contemporary solar flares, though far more powerful and associated with larger coronal structures. Although some differences are seen, MYStIX/SFiNCs super-flare behavior closely resembles solar long-duration events although with $10^4-10^7$ times more total energy. As in our study of super-flare frequency and energetics (Paper I), the modeling of flare processes provides no evidence that the flares arise in magnetic field lines connecting the star to the disk\footnote{A small fraction of flare-host stars, identified here as `diskless' using {\it Spitzer}-IRAC data, may possess anemic disks and accretion (\S \ref{sec:disk_dependence}).}. When solar flares and microflares are also considered, the similarity of flare physics spans an incredible $\sim 20$ orders of magnitude in energy. 

The paper is organized as follows.  Flare modeling of the bright MYStIX/SFiNCs super-flares, including evaluation of flaring coronal loop heights and widths, is presented in \S \ref{sec:methods2}. A summary of the COUP super-flares is given in the same section. Multi-variate analyses among the inferred flare and stellar properties for the bright MYStIX/SFiNCs super-flares are provided in \S \ref{sec:results2}. Comparison among the MYStIX/SFiNCs, COUP, and solar-stellar flares is considered in \S \ref{sec:comparison_among_flares}. Applicability of the solar single-loop model and dependencies of the PMS flare properties on stellar mass and disk are further discussed in \S \ref{sec:discussion2}. Appendices \ref{sec:photon_arrival} and \ref{sec:mle} provide details on the observed photon arrival diagrams of the 55 bright flares and address statistical issues in estimating flare physical parameters.

\begin{figure*}[ht!]
\epsscale{1.15}
\plotone{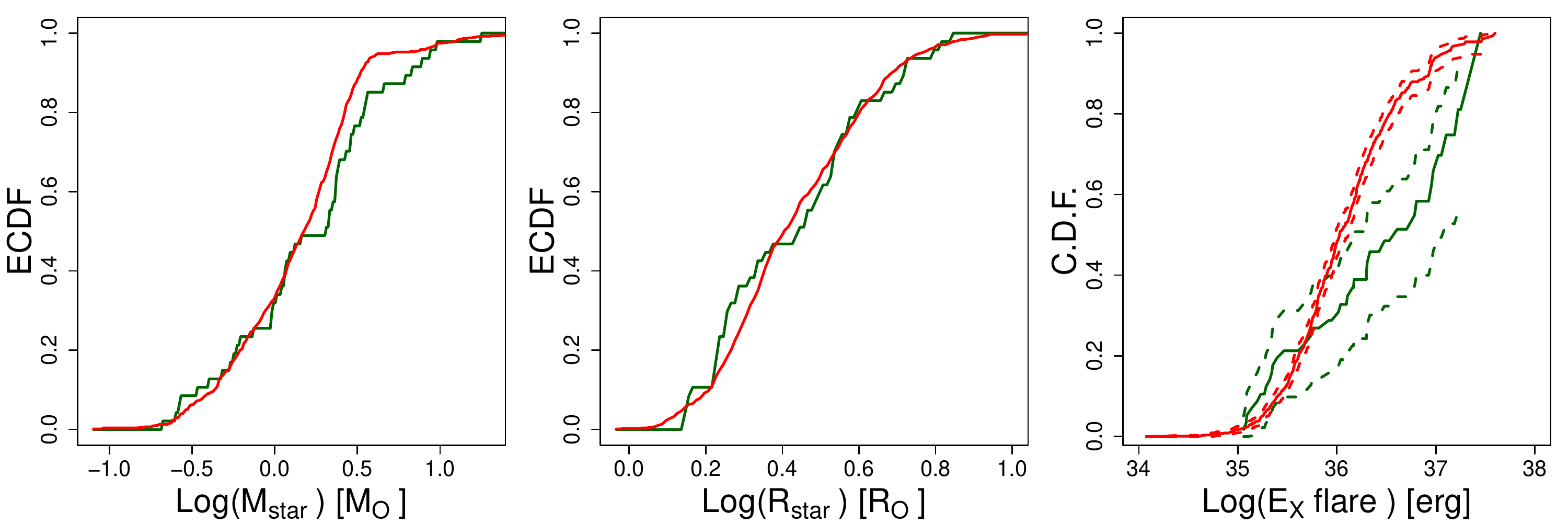}
\caption{Cumulative distribution functions of the X-ray flare host stellar mass (left), stellar radius (middle), and flare energy integrated from the start to the stop time (right). The latter is constructed using Kaplan-Meier survival estimators for `R' and `D' flares with the dashed curves representing 95\% confidence bands (see Paper I for details). The MYStIX/SFiNCs super-flares are stratified by brightness: the green curve shows the 55 events associated with $>1000$-photon {\it Chandra} observations (studied here) while the red curve shows the 1,031 remaining events associated with $<1000$-photon observations.  \label{fig:frd_flares}}
\end{figure*}

\section{Data and Modeling Procedure} \label{sec:methods2}

\subsection{Flare Sample} \label{sec:bright_flare_selection}

The selection of 1,086 super-flares from the {\it Chandra} MYStIX/SFiNCs observations is described in Paper I.  Super-flare candidates are first identified by two quantitative criteria: time-averaged X-ray luminosity exceeding $\log L_X =30.5$ erg~s$^{-1}$ and the probability of constant photon arrival times (based on a Kolmogorov-Smirnov test) $P_{KS} < 0.01$.  The 3,142 lightcurves satisfying these criteria are then analyzed with a Poisson multiple changepoint model similar to Bayesian Blocks \citep{Scargle2013}, and are then visually examined. The majority were found to have negligible or slow variations, but 1,086 had rapid flare-like variations. In 648 cases, the full (classified as `F') flare is seen with pre-flare and post-flare levels. In 438 cases, only the rise (`R') or decay (`D') portions of a flare are seen during the {\it Chandra} exposure.  Photons arriving between the start and stop time from the changepoint model, sometimes broadened slightly to catch the beginning of the rise and end of the decay phases,  are analyzed here.  Paper I provides a tabulation of flare and host star properties for the 1,086 super-flares. 

For astrophysical plasma modeling in the present study, we restrict consideration to about half of the brightest of the MYStIX/SFiNCs super-flares, i.e., only those detected in $>1000$-photon {\it Chandra} observations (see Appendix \ref{sec:photon_arrival} for details).  Fifty-five flares satisfy this criterion. Figure \ref{fig:frd_flares} compares three properties of this restricted and the full sample of super-flares.  These  some of the brightest X-ray flaring PMS stars (green curves) have masses and radii typical of the hosts of the full sample (based on nonparametric Anderson-Darling two-sample tests), but with flare energies representative of the most powerful events: two-thirds of these bright 55 flares have total energies in excess of $E_X = 1 \times 10^{36}$~erg. 

\subsection{Atlas of Flare Evolution} \label{sec:masme_method}

Following the methods of \citet{Getman08a}, adaptively smoothed moving medians of the time series of the X-ray counts and median energies are constructed for each of the 55 flare events. X-ray median energy is a surrogate for both absorbing column density and plasma temperature \citep{Getman10}; the bulk of absorption arises from the local environment of the star and should be constant in time\footnote{Some studies of individual flares do report increase in absorption during flares \citep[][]{Favata1999, Giardino2007b, Pillitteri2019, Moschou2019}.}, while the plasma temperature will change as flare plasma evolves. The median energy and count rate in adaptive intervals are converted to plasma temperature and intrinsic X-ray luminosity and emission measure employing calibrations from X-ray spectral simulations for a wide grid of temperatures with an assumption of single temperature optically thin thermal plasma subject to previously known flare-host's X-ray absorption \citep[\S 2.2 in][]{Getman08a}. 

This adaptive smoothing method neglects the contribution of the quiescent background X-ray emission or ``characteristic'' emission \citep{Wolk05,Caramazza07} that is probably the product of numerous superposed micro-flares and nano-flares. In their Appendix A, \citet{Getman08a} show that the contribution of the characteristic component to the emission of X-ray super-flares generally results in $<20$\% differences in the inferred peak plasma temperature and loop heights. It is interesting to note that the inspection of the electronic atlas of the COUP super-flares (Figure Set 2 in Getman et al. 2008a) and start/stop times for the COUP flare and characteristic emission segments (their Table~1) suggests that many PMS super-flares are immediately preceded/followed by an elevated X-ray emission for $\sim 4$~days. No correlation is seen between the duration of such elevated emission and stellar rotation period or radius; it is unclear if this elevated pre-flare emission is associated with the super-flare-host active region or other region(s).

\begin{figure*}[ht!]
\epsscale{1.15}
\plotone{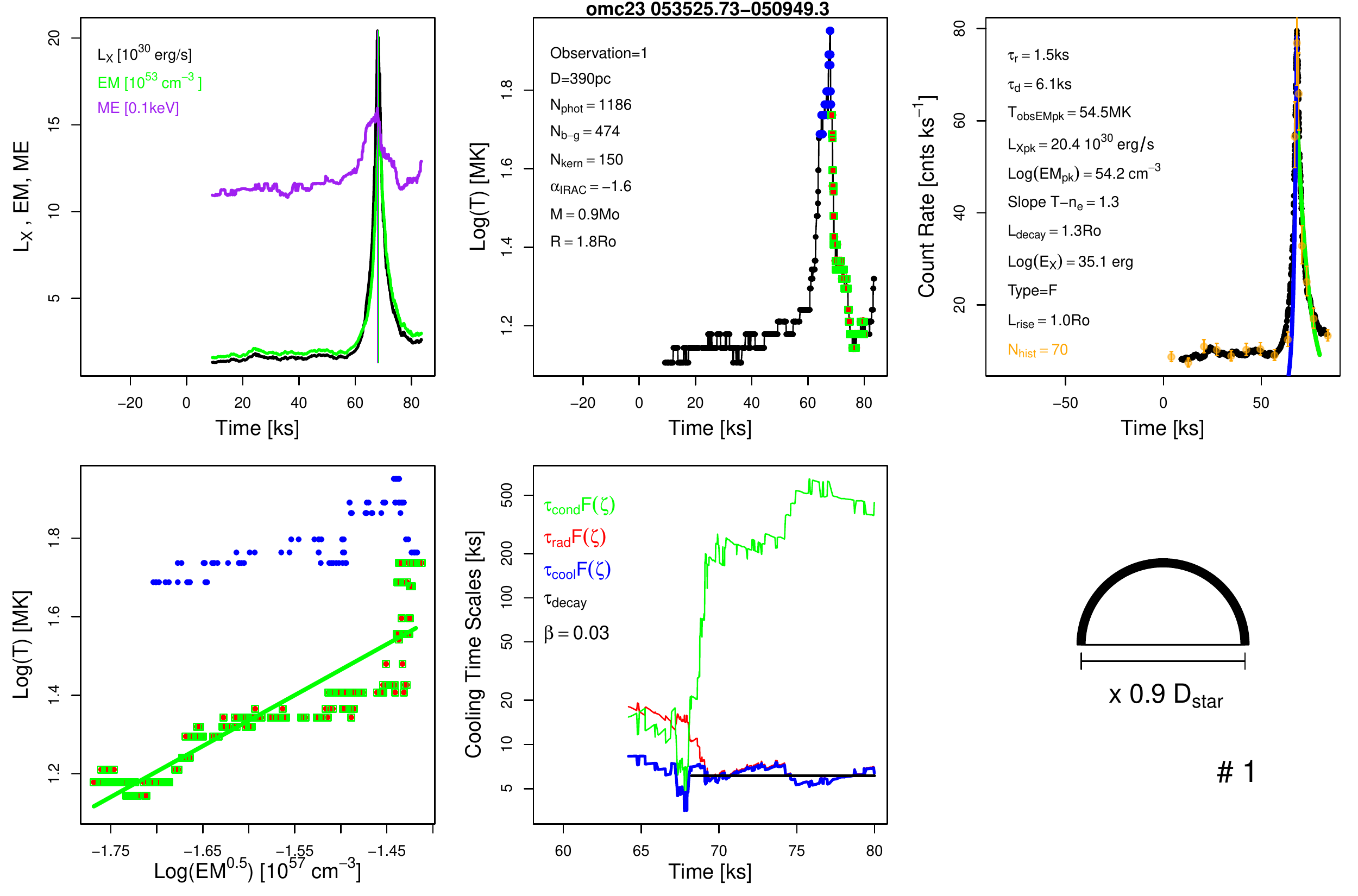}
\caption{Sample page from the MYStIX/SFiNCs super-flare atlas. See \S\ref{sec:masme_method} for explanations of the figures and legends.  The complete figure set for 55 super flares is available in the electronic article.  \label{fig:atlas_example1}}
\end{figure*}

Optimal kernel widths for our adaptively smoothed data are chosen based on the considerations described in Appendix \ref{sec:mle}. The MYStIX/SFiNCs kernel widths vary from $110$ to $600$ X-ray counts, similar to the kernel size range employed in the modeling of the COUP flares \citep{Getman08a}. Thus the statistical errors of $25-55$\% for plasma temperatures and $\la 0.2$~dex for log X-ray luminosity reported for the Orion super-flares should be applicable to the MYStIX/SFiNCs super-flares here.

Start (stop) time points for the modelled flare rise (decay) segments, with the point for the maximum smoothed count rate in-between, are initially chosen by visual inspection (with a concern about contamination from possible multiple re-heating events) and are later adjusted based on the choice of the optimal adaptive kernel size (Appendix \ref{sec:mle}). 

The rise ($\tau_{r}$) and decay ($\tau_d$) timescales for the flare lightcurves are derived using the maximum likelihood estimator on unbinned (raw) data as described in Appendix \ref{sec:mle}.

An example of the graphical output from these calculations is provided in Figure~\ref{fig:atlas_example1}. This example is a flare event originating in a member of the OMC-2/3 star forming region with SFiNCs designation 053525.73-050949.3 which is the lightly-absorbed classical T-Tauri star AH~Ori. 

The top-left panel shows the evolution of three properties: X-ray luminosity (black curve), emission measure (green), and median energy (magenta).  The vertical lines mark the peaks of the median energy and emission measure. The top middle and right panels show the temporal evolution of the flare plasma temperature and observed count rate.  Here the blue curve shows the best-fit exponential fit to the flare rise and the green curve is the fit to the flare decay. These are based on the fits to the unbinned data (Appendix \ref{sec:mle}). For comparison, the count rates and their 1-$\sigma$ confidence bands in independent bins are shown as orange points with error bars.

The upper center and right panels of the atlas also give a variety of scalar properties of the host star and the X-ray flare. Properties taken from Paper~I include: the sequence number for multiple ObsID observations that captured the flare; distance from the Sun; number of X-ray counts in the related {\it Chandra} ObsID ($N_{phot}$); number of X-ray counts within the rise-decay phases (blue-green) employed in our flare characterization here ($N_{b-g}$); number of photons used in the smoothing kernel; stellar mass; stellar radius; and {\it Spitzer}-IRAC spectral energy distribution (SED) slope. ``NaN'' is listed when property value is not available. Flare characteristics estimated in the current work include: e-folding rise and decay timescales for unbinned data (Appendix \ref{sec:mle}); peak flare X-ray luminosity; peak emission measure\footnote{Emission measure is the integral of density squared over the loop volume. Here we assume the density is uniform over this volume which is not likely to be accurate.}; temperature at the time of peak emission measure;  flare energy calculated as the integral of $L_X$ from the flare's start to stop times\footnote{An alternative energy quantity ($L_X \times \tau_{decay}$) is employed below  for consistent comparison with the flare energies released by Orion young stars (\S \ref{sec:coup2}).  The two quantities generally differ by $<60$\%.}; and flare type ``F'' or ``R'' or ``D''.  The slope $\zeta$ is obtained from the decay phase in the log-temperature $vs.$ log-density diagram for smoothed data shown in the lower-left panel.  Two estimates of flaring loop heights from the decay and rise phases are derived from the flare model in \S\ref{sec:flare_model}. The size of the independent count rate bins, in X-ray counts, is also listed (in orange).

The bottom left panel of the atlas page shows the evolution of temperature as a function of the square root of the emission measure that scales with plasma density. This is the diagram where the four phases of heating and cooling from the \citet{Reale1997} model can be traced.  Blue points represent the rise phase, red points (if present) represent the segment between the peaks of the count rate and emission measure, and green points indicate the decay phase after the peak of the emission measure. In 76\% of the bright MYStIX/SFiNCs super-flares, the temperature peak clearly precedes the emission measure peak, although this is not a strong effect in the example of Figure~\ref{fig:atlas_example1}. The green line shows a least squares linear regression of the decay phase.  The slope $\zeta$ is measured for the trajectory after the emission measure peak. 

The bottom middle and right panels are described in \S\ref{sec:cooling}. The bottom right corner gives the running number of super-flares, 1 through 55. 
  
Table~\ref{tab:55_brightest} lists properties for the 55 bright MYStIX/SFiNCs super-flares and their stellar hosts. The derivation of these properties in Table~\ref{tab:55_brightest} is outlined in next sections. The reported quantities include flare host stellar mass, radius, and SED IRAC slope; flare type (`F' or `R' or `D'); rise and decay X-ray light-curve timescales; flare peak X-ray luminosity, emission measure, plasma temperature (including  temperature at the moment of peak emission measure); inferred flaring coronal loop heights during rise and decay, and loop width ratio $\beta$.  The super-flares are listed in order of increasing $\beta$, when known.   

\startlongtable
\begin{deluxetable*}{lcccccrrrrrrcrr}
\tabletypesize{\scriptsize}
\tablecaption{Fifty Five Bright MYStIX/SFiNCs X-ray Super-flares \label{tab:55_brightest}}
\tablewidth{0pt}
\tablehead{
~ \\
\multicolumn{4}{c}{Host Star Properties} && \multicolumn{10}{c}{Super-flare Properties} \\ \cline{1-4} \cline{6-15} 
\colhead{MYStIX/SFiNCs} & \colhead{$M$} &
\colhead{$R$} & \colhead{$\alpha_{IRAC}$} && \colhead{Type} & \colhead{$\tau_{rise}$} & \colhead{$\tau_{decay}$} & \colhead{$L_{X,pk}$} & \colhead{$EM_{pk}$} & \colhead{$T_{obs,pk}$} &  \colhead{$T_{obs,EM}$} & \colhead{$L_{decay}$} & \colhead{$L_{rise}$} & \colhead{$\beta$} \\ 
\colhead{} & \colhead{$M_{\odot}$} & \colhead{$R_{\odot}$} & \colhead{} && \colhead{} & \colhead{ks} & \colhead{ks} & 
\colhead{${10^{30}}$} & \colhead{$10^{54}$} & \colhead{MK} & \colhead{MK} & \colhead{$R_{\odot}$} & \colhead{$R_{\odot}$} & \colhead{} \\
\colhead{} & \colhead{} & \colhead{} & \colhead{} && \colhead{} & \colhead{} &
\colhead{} & \colhead{erg/s} & \colhead{cm$^{-3}$} &
\colhead{} & \colhead{} & \colhead{} &
\colhead{} & \colhead{} \\
\colhead{(1)} & \colhead{(2)} & \colhead{(3)} & \colhead{(4)} && \colhead{(5)} & \colhead{(6)} &
\colhead{(7)} & \colhead{(8)} & \colhead{(9)} &
\colhead{(10)} & \colhead{(11)} & \colhead{(12)} &
\colhead{(13)} & \colhead{(14)}
}
\decimals
\startdata
1 053525.73-050949.3\_1 & 0.9 & 1.8 & -1.6 && F & $1.5 \pm 0.3$ & $6.1 \pm 1.5$ & 20 & 1.5 & 89 & 55 & 1.3 & 1.0 & 0.03\\
2 053515.79-053312.0\_1 & 0.3 & 1.8 & -1.4 && F & $6.6 \pm 2.3$ & $30.3 \pm 11.0$ & 11 & 1.3 & 31 & 22 & 2.4 & 1.7 & 0.05\\
3 053515.91-051458.8\_1 & 0.3 & 2.1 & -2.3 && F & $4.1 \pm 0.8$ & $10.9 \pm 1.1$ & 80 & 6.7 & 59 & 38 & 1.7 & 1.9 & 0.06\\
4 034359.68+321402.7\_1 & 0.5 & 1.5 & -2.4 && F & $1.5 \pm 0.2$ & $9.9 \pm 1.7$ & 37 & 3.5 & 53 & 28 & 1.5 & 1.1 & 0.07\\
5 053447.62-054350.7\_1 & 0.6 & 1.7 & -2.9 && F & $13.5 \pm 11.8$ & $20.0 \pm 14.3$ & 8 & 0.9 & 21 & 17 & 1.4 & 1.8 & 0.08\\
6 054703.96+001114.0\_1 & 0.5 & 1.8 & -1.2 && F & $6.7 \pm 1.5$ & $12.3 \pm 1.2$ & 43 & 4.0 & 136 & 29 & 1.4 & \nodata & 0.09\\
7 034432.73+320837.3\_2 & 1.2 & 2.1 & -2.4 && F & $6.0 \pm 1.7$ & $6.8 \pm 2.5$ & 14 & 1.5 & 29 & 23 & 0.6 & 1.1 & 0.11\\
8 064058.51+093331.7\_1 & 2.3 & 2.4 & -2.9 && F & $2.0 \pm 0.5$ & $20.0 \pm 2.0$ & 113 & 9.5 & 99 & 27 & 1.5 & 10.2 & 0.10\\
9 064056.50+095410.4\_1 & \nodata & \nodata & -2.9 && D & $3.6 \pm 3.4$ & $19.2 \pm 2.0$ & 74 & 7.4 & 36 & 26 & 1.5 & \nodata & 0.10\\
10 054635.36+000858.4\_1 & 0.2 & 1.9 & -2.8 && F & $4.2 \pm 1.9$ & $8.9 \pm 4.5$ & 17 & 2.0 & 22 & 20 & 1.0 & 0.5 & 0.12\\
11 053523.44-051051.7\_1 & 1.1 & 1.9 & -2.2 && F & $11.0 \pm 5.9$ & $18.5 \pm 9.7$ & 4 & 0.5 & 15 & 13 & 1.3 & 1.2 & 0.13\\
12 053526.29-050840.0\_1 & \nodata & \nodata & -2.8 && F & $2.0 \pm 0.8$ & $4.1 \pm 2.1$ & 13 & 1.5 & 21 & 20 & 0.3 & 0.2 & 0.19\\
13 071845.25-245643.9\_1 & 0.4 & 1.4 & -2.7 && F & $2.7 \pm 0.5$ & $7.1 \pm 4.2$ & 195 & 18.3 & 50 & 29 & 1.2 & 1.5 & 0.17\\
14 053456.81-051133.1\_1 & 1.0 & 1.7 & -1.6 && F & $2.6 \pm 0.5$ & $7.9 \pm 1.3$ & 15 & 1.7 & 17 & 17 & 0.4 & 0.2 & 0.19\\
15 085932.21-434602.3\_1 & 2.5 & 3.9 & -2.6 && F & $1.9 \pm 0.2$ & $5.3 \pm 0.4$ & 818 & 67.5 & 415 & 45 & 0.8 & \nodata & 0.19\\
16 060814.04-062559.5\_1 & 9.5 & 3.8 & -0.6 && F & $3.4 \pm 0.9$ & $3.7 \pm 0.7$ & 514 & 40.2 & 65 & 45 & 0.2 & 1.4 & 0.29\\
17 054145.08-015144.4\_1 & 3.0 & 5.1 & -2.0 && F & $5.8 \pm 3.4$ & $13.9 \pm 3.9$ & 53 & 4.1 & 82 & 51 & 2.4 & 3.6 & 0.30\\
18 180402.88-242140.0\_3 & 2.9 & 4.6 & -2.6 && F & $2.3 \pm 0.6$ & $7.3 \pm 1.1$ & 176 & 11.5 & 178 & 66 & 1.9 & 7.8 & 0.33\\
19 043040.90+351352.3\_1 & \nodata & \nodata & -0.2 && F & $3.5 \pm 0.9$ & $5.1 \pm 1.3$ & 92 & 6.9 & 501 & 53 & 1.0 & \nodata & 0.48\\
20 034435.37+321004.5\_1 & 3.3 & 5.3 & -1.8 && R & $35.7 \pm 25.2$ & $34.5 \pm 38.7$ & 12 & 1.3 & 26 & 20 & 4.1 & 6.5 & 0.55\\
21 064105.36+093313.5\_1 & 1.2 & 1.8 & -2.6 && F & $5.4 \pm 1.2$ & $21.3 \pm 3.2$ & 57 & 4.9 & 63 & 35 & 3.9 & 3.7 & 0.59\\
22 053453.03-050327.0\_1 & 0.9 & 1.4 & -1.2 && F & $7.0 \pm 1.9$ & $8.4 \pm 4.4$ & 7 & 0.7 & 52 & 22 & 1.4 & 8.4 & 0.64\\
23 053531.95-050927.9\_1 & 8.7 & 3.6 & -0.9 && F & $7.0 \pm 2.2$ & $16.7 \pm 6.8$ & 8 & 0.9 & 30 & 23 & 2.1 & 1.4 & 0.62\\
24 023308.87+612522.2\_1 & 2.0 & 3.3 & -2.8 && F & $3.4 \pm 0.5$ & $8.5 \pm 1.8$ & 1693 & 157.2 & 45 & 28 & 0.6 & 1.5 & 0.67\\
25 034427.01+320443.8\_3 & 0.6 & 1.4 & -2.2 && F & $1.1 \pm 0.1$ & $5.3 \pm 1.7$ & 81 & 6.0 & 555 & 656 & \nodata & 0.5 & \nodata\\
26 225655.34+624223.9\_1 & 2.4 & 3.8 & -2.8 && F & $0.6 \pm 0.1$ & $10.3 \pm 1.8$ & 697 & 49.9 & 655 & 655 & \nodata & 0.4 & \nodata\\
27 034450.60+321905.9\_1 & 6.7 & 3.1 & -2.4 && R & $5.2 \pm 1.9$ & $4.3 \pm 1.1$ & 16 & 1.6 & 30 & 26 & \nodata & 0.8 & \nodata\\
28 053516.75-044032.3\_1 & 1.4 & 2.7 & \nodata && F & $1.5 \pm 0.2$ & $12.2 \pm 3.9$ & 25 & 2.4 & 51 & 31 & \nodata & 0.7 & \nodata\\
29 182041.13-161530.8\_2 & \nodata & \nodata & -2.0 && F & $1.2 \pm 0.2$ & $16.9 \pm 1.7$ & 1325 & 108.8 & 57 & 39 & \nodata & 0.5 & \nodata\\
30 182028.36-161030.6\_2 & 2.9 & 4.9 & \nodata && R & $10.3 \pm 12.4$ & $10.3 \pm 5.4$ & 376 & 30.3 & 51 & 45 & \nodata & 2.1 & \nodata\\
31 054131.61-015231.7\_1 & 0.2 & 1.4 & -0.6 && F & $9.1 \pm 12.4$ & $13.5 \pm 6.2$ & 49 & 4.3 & 43 & 26 & \nodata & 4.3 & \nodata\\
32 053536.67-050414.3\_1 & 0.3 & 1.7 & -1.8 && F & $2.5 \pm 0.4$ & $50.0 \pm 53.2$ & 13 & 1.0 & 64 & 38 & \nodata & 1.5 & \nodata\\
33 225355.16+624337.0\_1 & 2.4 & 3.4 & -1.9 && F & \nodata & $9.0 \pm 4.8$ & 432 & 39.2 & 50 & 34 & \nodata & 5.0 & \nodata\\
34 064040.44+095050.4\_1 & \nodata & \nodata & \nodata && R & $13.5 \pm 3.7$ & $83.3 \pm 176.4$ & 72 & 7.0 & 43 & 26 & \nodata & 6.3 & \nodata\\
35 054134.34-020150.9\_1 & 1.1 & 1.7 & -2.8 && F & $40.0 \pm 27.4$ & $55.6 \pm 380.2$ & 4 & 0.5 & 15 & 12 & \nodata & 5.4 & \nodata\\
36 180438.93-242533.2\_1 & 0.7 & 1.7 & -2.9 && F & $0.4 \pm 0.0$ & $4.3 \pm 0.4$ & 1304 & 125.7 & 91 & 20 & \nodata & 3.3 & \nodata\\
37 180358.83-242529.2\_1 & 2.1 & 3.2 & -2.7 && R & $7.6 \pm 0.9$ & $20.4 \pm 96.0$ & 2091 & 181.2 & 81 & 36 & \nodata & 10.6 & \nodata\\
38 054138.24-015309.1\_1 & 3.4 & 6.3 & -2.2 && F & $5.2 \pm 2.3$ & $16.1 \pm 8.6$ & 48 & 3.7 & 95 & 46 & \nodata & 6.3 & \nodata\\
39 053528.28-045838.4\_1 & \nodata & \nodata & 0.2 && F & $6.3 \pm 1.2$ & $40.0 \pm 16.7$ & 33 & 2.9 & 63 & 30 & \nodata & 6.2 & \nodata\\
40 183011.15+011238.2\_1 & \nodata & \nodata & -0.8 && R & $6.8 \pm 1.1$ & $52.6 \pm 14.9$ & 32 & 3.1 & 49 & 19 & \nodata & 10.0 & \nodata\\
41 034444.60+320402.7\_3 & 2.1 & 2.8 & -0.9 && R & $3.6 \pm 0.5$ & $33.3 \pm 5.8$ & 250 & 25.9 & 77 & 21 & \nodata & 15.5 & \nodata\\
42 054643.37+000436.0\_1 & 4.5 & 3.0 & -2.8 && R & $8.4 \pm 1.8$ & $111.1 \pm 135.6$ & 69 & 6.7 & 77 & 27 & \nodata & 20.1 & \nodata\\
43 054707.92+001756.1\_1 & 17.6 & 5.3 & -2.8 && D & $8.5 \pm 4.9$ & $29.4 \pm 8.6$ & 64 & 6.2 & 78 & 28 & \nodata & 19.3 & \nodata\\
44 182031.11-160929.8\_2 & \nodata & \nodata & -2.7 && R & $23.8 \pm 14.2$ & $25.0 \pm 12.8$ & 229 & 16.3 & 100 & 51 & \nodata & 26.6 & \nodata\\
45 054148.21-015601.9\_1 & \nodata & \nodata & 0.1 && D & \nodata & $14.7 \pm 1.6$ & 120 & 8.1 & 380 & 94 & \nodata & \nodata & \nodata\\
46 060747.99-062537.7\_1 & 1.3 & 3.5 & 0.2 && D & \nodata & $41.7 \pm 4.7$ & 283 & 21.0 & 68 & 31 & \nodata & \nodata & \nodata\\
47 180412.48-241943.2\_4 & 2.2 & 2.9 & -2.9 && D & $1.3 \pm 1.0$ & $11.0 \pm 1.8$ & 337 & 26.4 & 64 & 46 & \nodata & \nodata & \nodata\\
48 223940.26+751321.6\_1 & 6.1 & 2.9 & -0.0 && D & \nodata & $45.5 \pm 15.4$ & 17 & 1.6 & 34 & 28 & \nodata & \nodata & \nodata\\
49 183127.65-020509.6\_1 & 3.6 & 6.5 & -2.4 && D & \nodata & $7.1 \pm 0.7$ & 245 & 21.3 & 638 & 27 & \nodata & \nodata & \nodata\\
50 053513.00-053934.8\_1 & 7.8 & 3.4 & -1.3 && D & \nodata & $58.8 \pm 12.7$ & 24 & 2.5 & 31 & 23 & \nodata & \nodata & \nodata\\
51 225504.09+623418.9\_3 & 0.6 & 1.7 & -1.3 && D & \nodata & $14.1 \pm 2.6$ & 191 & 17.5 & 57 & 27 & \nodata & \nodata & \nodata\\
52 182039.04-160836.9\_2 & 2.7 & 4.0 & \nodata && D & \nodata & $24.4 \pm 3.7$ & 515 & 40.7 & 55 & 42 & \nodata & \nodata & \nodata\\
53 111105.63-610146.1\_1 & 3.6 & 7.0 & -2.7 && D & \nodata & $23.8 \pm 3.0$ & 1007 & 118.8 & 22 & 19 & \nodata & \nodata & \nodata\\
54 225355.16+624337.0\_2 & 2.4 & 3.4 & -1.9 && F & $9.0 \pm 14.8$ & $2.6 \pm 1.9$ & 513 & 41.4 & 86 & 41 & \nodata & \nodata & \nodata\\
55 034437.87+320804.1\_1 & 1.1 & 2.2 & -1.0 && D & \nodata & $34.5 \pm 4.5$ & 27 & 2.4 & 63 & 24 & \nodata & \nodata & \nodata\\
\enddata
\tablecomments{Column 1: Flare name composed of the host star name and the relative number of the {\it Chandra} X-ray observation, during which the flare was detected (see Paper~I for details). Columns 2-4: Stellar mass, radius, and infarred spectral energy distribution slope from Paper~I. Values of $\alpha_{IRAC} < -1.9$ and $\alpha_{IRAC} > -1.9$ indicate diskless and disky stars, respectively. Column 5: Flare type: `F' - full or nearly full flare was captured; `R' - mainly the flare rise phase was captured; `D' - mainly the flare decay phase was captured. Columns 6-7: Flare rise and decay time-scales and their 95\% CIs derived using MLE estimator on unbinned data (\S \ref{sec:mle}).} Columns 8-11: Flare peak X-ray luminosity (in units of $10^{30}$~erg~s$^{-1}$), peak emission measure (in units of $10^{54}$~cm$^{-3}$), peak plasma temperature, and plasma temperature at an instant of peak emission measure. For the following `R'- and `D'-type flares, these peak values are likely lower limits to the true values: \#37, 45, 46, 47, 48, 49, 50, 51, 53, and 55. Columns 12-13: Heights of the flaring coronal loops inferred from the two different methods described in the text.  Column 14: The ratio between the loop’s cross-sectional radius and the loop height.
\end{deluxetable*}

\begin{deluxetable*}{rccrrrrrrrc}
\tabletypesize{\small}
\tablecaption{Two Hundred Sixteen COUP X-ray Super-flares \label{tab:216_coup}}
\tablewidth{0pt}
\tablehead{
\colhead{Name} & \colhead{$M$} &
\colhead{$R$} & \colhead{$\alpha_{IRAC}$} & \colhead{$\tau_{rise}$} & \colhead{$\tau_{decay}$} & \colhead{$L_{X,pk}$} & \colhead{$EM_{pk}$} & \colhead{$T_{obs,pk}$} &  \colhead{$T_{obs,EM}$} & \colhead{$L_{decay}$}\\
\colhead{} & \colhead{$M_{\odot}$} & \colhead{$R_{\odot}$} & \colhead{} & \colhead{ks} & \colhead{ks} & \colhead{} & \colhead{} & \colhead{MK} & \colhead{MK} & \colhead{$R_{\odot}$}\\
\colhead{(1)} & \colhead{(2)} & \colhead{(3)} & \colhead{(4)} & \colhead{(5)} & \colhead{(6)} &
\colhead{(7)} & \colhead{(8)} & \colhead{(9)} &
\colhead{(10)} & \colhead{(11)}
}
\startdata
7\_1~~~ & \nodata & \nodata & \nodata~~ & 13.5 & 13.0~~ & 42~~~ & 4.0~~~ & 41~~~~ & 28~~~~ & 1.9\\
9\_1~~~ & 1.8 & 2.6 & \nodata~~ & 10.0 & 19.2~~ & 151~~~ & 11.7~~~ & 103~~~~ & 32~~~~ & \nodata\\
11\_1~~~ & 2.7 & 4.2 & -1.3~~~~ & 3.4 & 6.5~~ & 69~~~ & 4.8~~~ & 696~~~~ & 696~~~~ & \nodata\\
23\_1~~~ & 2.4 & 3.4 & -2.7~~~~ & 2.4 & 5.5~~ & 96~~~ & 7.3~~~ & 60~~~~ & 55~~~~ & 1.6\\
27\_1~~~ & 0.7 & 1.4 & -2.6~~~~ & 15.4 & 27.9~~ & 4~~~ & 0.3~~~ & 41~~~~ & 31~~~~ & 5.2\\
\enddata
\tablecomments{Column 1: Name, composed of the COUP source number and flare number given in Tables 1-3 of \citet{Getman08a}. Columns 2-4: Stellar mass, radius, and infarred spectral energy distribution slope from Paper~I. Columns 7-10: Flare peak X-ray luminosity (in units of $10^{30}$~erg~s$^{-1}$), peak emission measure (in units of $10^{54}$~cm$^{-3}$), peak plasma temperature, and plasma temperature at an instant of peak emission measure. Column 11: Height of the flaring coronal loops. This table is available in its entirety (216 flares) in the machine-readable form in the online journal. A portion is shown here for guidance regarding its form and content.}
\end{deluxetable*}

\subsection{Loop Height} \label{sec:flare_model}

Having detailed data on the decay of emission measure and temperature during the flare cooling phase, we can infer the length of a flaring coronal structure within the framework of the solar flare model of \citet{Reale1997}. This is a 1-dimensional  time-dependent hydrodynamic model of a semi-circular coronal magnetic loop with  constant cross-section uniformly filled with plasma.  This plasma is subject to instantaneous or prolonged heating, viscosity, radiative losses, and thermal conduction in the gravitational field of the host star. 

The hydrodynamical simulations of \citet{Reale1997} provide a  relation between loop height and three observable parameters for the flare decay phase: the flare decay exponential timescale $\tau_{decay}$; the plasma temperature at the instant of peak emission measure $T_{EM,pk}$; and the slope on the plasma log-temperature $vs.$ log-density diagram $\zeta$. The half-length $L_{decay}$ of a coronal loop is 
\begin{equation} \label{eqn:loop_decay}
L_{decay} = \frac{\tau_{decay} ~ (T_{EM,pk})^{1/2}} {\alpha ~ F(\zeta)}
\end{equation}
where $\alpha$ is a constant and $F(\zeta)$ is function that accounts for prolonged heating. Higher $\zeta$ values correspond to freely decaying loops with no sustained heating, while lower values correspond to loops with sustained heating. 

In their study of Orion Nebula Cluster flares seen in the 13-day COUP observation, \citet{Favata2005} calibrated the $\alpha$, $T_{EM,pk}$, $F(\zeta)$ quantities for the {\it Chandra}-ACIS detector as (assuming c.g.s. units):
\begin{eqnarray}
\alpha ~&=&~ 3.7 \times 10^{-4}  \nonumber \\
F(\zeta) ~&=&~ 0.63 / (\zeta - 0.32) + 1.41 \\
T_{EM,pk} ~&=&~ 0.068 ~  T_{obs,EM,pk}^{1.2} \nonumber
\end{eqnarray} 
where $T_{obs,EM,pk}$ is the plasma temperature at emission measure peak obtained in \S \ref{sec:masme_method}. The $F(\zeta)$ formula is applicable in the range $0.32 < \zeta \lesssim 1.5$. In our loop geometry calculations, we include two flares with $1.5 < \zeta < 1.8$. A few MYStIX/SFiNCs super-flares (\#25, 26, 28, 32, and 44) exhibit extremely steep slopes on the temperature-density diagram, $\zeta > 2.5$. Favata et al. suggest that for such flares ``the analyzed part of the decay is too short to cover a significant part of the decay path, and the slope is not yet well defined''. Finally, a  number of MYStIX/SFiNCs super-flares have near 0 or negative $\zeta$ due to either incomplete coverage of their flare decay phases (\#27, 30, 34, 37, 39, 40, 41, 42) and/or complex multiple re-heating events (\#29, 31, 36, 43, 54, 55). For these events, we do not estimate $L_{decay}$. 

\citet{Reale2007} further calibrate this hydrodynamic flaring loop model to the temporal evolution of plasma temperature during the rise phase of the flare, providing an alternative formula for the loop half-length $L_{rise}$.  The relation is (in c.g.s. units)   
\begin{equation} \label{eqn:loop_rise}
L_{rise} = 950 ~ \sqrt{T_{max}} ~\tau_{rise} ~ \Psi^2
\end{equation} 
where $T_{max}$ is the maximum plasma temperature and $\Psi = T_{max}/T_{EM,pk}$. For the remainder of this study, the term ``loop height''  will be used to indicate the loop half-length in equation (\ref{eqn:loop_decay}) or equation (\ref{eqn:loop_rise}). 

The typical errors on peak plasma temperature of $25\%-55$\% result in $15\%-30$\% errors on inferred loop heights. For seventeen flares with both $L_{decay}$ and $L_{rise}$ estimates available, there is a strong correlation between the two inferred alternative loop heights with no biases. For three flares (\#\# 8, 18, 22), $L_{rise}$ appears to be systematically higher (by factors $4-7$) than $L_{decay}$. Since the factor $\Psi^2$ in equation \ref{eqn:loop_rise} is especially vulnerable to possible noise in the plasma temperature temporal evolution, we prefer $L_{decay}$ over $L_{rise}$ estimates.

We are able to estimate $L_{rise}$ for 39 of the 55 bright MYStIX/SFiNCs super-flares, and estimate $L_{decay}$ for 24 super-flares. $L_{rise}$ estimates are missing for type D flares where only the decay phase is seen, and $L_{decay}$ estimates are missing for type R flares or when $\zeta > 2$. In the super-flare atlas, $L_{rise}$ and $L_{decay}$ values are given in the upper right panel.  

Finally, we must consider the possibility that the uniform single-loop flare model of \citet{Reale1997} has too simple a geometry, or too simple pattern of heating input, for PMS super-flares.  Solar flares are often associated with arcades of multiple loops.  Reale et al.\  validated their model on 20 single-loop solar M- and C-class flares that were imaged directly by {\it Yohkoh}-SXT telescope.  \citet[][Appendix]{Getman2011} further applied the Reale et al. method to 5 solar X-class limb flares associated with arcades of multiple loops seen in EUV {\it TRACE} images. Surprisingly, in contrast to the general view that single-loop models would overestimate flaring loop heights of complex flare events,  they found that the model length predictions were comparable to or shorter than the lengths of the individual loops measured from the TRACE images.

\citet{Reale2004} argue that the single-loop modeling procedure can be applied to multi-loop arcades assuming the presence of a dominant loop. \citet{Getman2011} suggest that the single-loop method can provide reasonable loop height estimates in cases of multi-loop structures where individual loop events having similar temporal temperature and emission measure profiles and firing nearly simultaneously. The MYStIX/SFiNCs super-flares mostly have long duration with unimodal structure (Paper I) that suggests nearly simultaneous firing if multiple loops are involved.  Multiple individual loops firing at non-simultaneously would produce multimodal lightcurves with shorter sub-flare components  \citep{Aschwanden2001}. 

Altogether, we consider the loop heights from the \citet{Reale1997} model to be valid for MYStIX/SFiNCs superflares, though with substantial inaccuracies.

\subsection{Cooling Processes and Loop Thickness} \label{sec:cooling}

The \citet{Reale1997} model assumes a 1-D flaring loop with a constant cross section. In many solar and stellar applications, a single loop with a 10:1 ratio between the loop height and cross-sectional radius $r$ is assumed; that is 
\begin{equation}
\beta ~=~ r/L   
\end{equation}
is often fixed at $\beta=0.1$. This value was assumed in the earlier studies of solar flares by \citet{Reale1997} and Orion super-flares by \citet{Favata2005} and \citet{Getman08a}. In the  \citet{Reale1997} model, the slope on the temperature-density diagram and the loop length do not depend on this parameter $\beta$.

However, as shown by \citet{Getman2011}, the loop thickness ratio $\beta$ can be estimated from the data within the framework of  \citet{Reale1997} model. Thermal conduction and radiation timescales can be expressed in terms of the $\beta$ and other available quantities, such as the loop height $L$ and temporal profiles of temperature $T(t)$, X-ray luminosity $L_X(t)$, and emission measure $EM(t)$ according to
\begin{equation} \label{eq:cooling_timescales}
\tau_{con}(t) = \frac{3 k_b}{\kappa_0 T(t)^{5/2} \beta} \sqrt{\frac{EM(t) L}{2 \pi}} ,
\end{equation}
\begin{equation} \label{eq:radiation_timescales}
\tau_{rad}(t) = \frac{3 k_b T(t) \beta}{L_{X}(t)} \sqrt{EM(t) 2 \pi L^3} ,
\end{equation}
where $k_b$ is the Boltzmann constant and $\kappa_0$ is the coefficient of thermal conductivity.  The cooling time that combines both conduction and radiation processes is $\tau_{th}$ where 
\begin{equation}
 1/\tau_{th} = 1/\tau_{con} + 1/\tau_{rad} 
\end{equation}
The value of $\beta$ that gives $\tau_{th} $ closest to the observed flare decay timescale (corrected for possible sustained heating), $\tau_{decay} / F(\zeta)$, is most consistent with the observed X-ray flare properties. It is important to note that, as in \citet{Getman2011}, the above $L_X$ quantity in equation \ref{eq:radiation_timescales} is calculated in the wide energy band $[0.01 - 50]$~keV.  

The lower-middle panel in Figure \ref{fig:atlas_example1} shows an example of the inferred temporal profiles of the cooling timescales for thermal conduction (green curve), radiation (red), combined conduction plus radiation (blue), and the exponential fit to the observed lightcurve ($\tau_{decay} = 6.1$~ks; black line). In this case, a good fit is seen between the observed (black) and model (blue) cooling timescales.   The analysis yields a loop thickness $\beta = 0.03$, shown in the figure legend.  In this case, the loop thickness is consistent with solar coronal loops where $\beta$ ranges from $\la 0.01$ to 0.1 in X-ray images \citep{Handy1999,Aschwanden2000}. 

The lower-right panel in the atlas shows a 2-dimensional schematic of the flaring coronal structure with a semi-circular loop with height $L_{decay}$ and width ratio $\beta$.  The length of the structure's base is indicated in units of stellar diameter.

The loop thickness parameter $\beta$ is used to sequence the flares in the atlas associated with Figure \ref{fig:atlas_example1} provided in the electronic edition.  The flares are placed in order of increasing flaring loop thickness from $\beta = 0.03$ to $\beta \geq 0.6$.

\subsection{Bootstrap Analysis} \label{sec:error_analysis}
For each of the 24 modelled flares with available loop geometries (Table \ref{tab:55_brightest}), a hundred re-samples of original flare X-ray photons with replacements are generated.  These bootstrapped datasets are passed through our flare analysis code to produce distributions of flare peak temperature, X-ray luminosity and emission measure, temperature-density slope, loop height and width. Normalized median absolute deviations around medians of such simulated flare property distributions are considered here as measures of formal statistical errors. 

We find that typical errors on peak temperatures and X-ray luminosities are around 20\% and 10\%, respectively. Errors on temperature-density slopes and loop geometries often reach $50-100$\%, but the $\zeta$ and $L_{decay}$ properties derived in our flare modelling and reported in the flare atlas (Figure Set \ref{fig:atlas_example1}) and Table \ref{tab:55_brightest} are generally lie within 15\% of the simulated distribution medians. Simulated medians on the loop thickness parameter $\beta$ are $\lesssim 0.1$ for the first 11 flares in Table~\ref{tab:55_brightest} and $\gtrsim 0.2$ for the remaining of the 24 flares. 

These uncertainties in derived physical flare parameters arise primarily from the low count rates from MYStIX/SFiNCs super-flares due to their $\sim 1-2$~kiloparsec distance.  Nonetheless, in our science analyses below, we are not so much interested in exact loop geometries and plasma properties for individual flares as in trends among flare and star properties for ensembles of flares. These trends should not be seriously affected by the statistical noise of the fits to the \citet{Reale1997} model with variable $\beta$.

\subsection{COUP Super-Flares} \label{sec:COUP_superflares_description}

In \S\ref{sec:coup2}, these modeling results for the 55 bright MYStIX/SFiNCs super-flares will be compared to the properties of the previously published Orion Nebula Cluster X-ray super-flares \citep{Getman08a,Getman08b}. The MYStIX/SFiNCs and COUP flares have similar appearances, and Paper~I shows that the power law slope of the super-flare energy distribution, $dN/dE_{X} \propto E_X^{-2}$, are consistent for the two samples as well as for older stars and the Sun.  

Getman et al. analyzed 216 X-ray brightest super-flares from 161 young stellar objects observed during the nearly continuous $\sim 13$ day COUP exposure of the Orion Nebula \citep{Getman05}, the deepest continuous exposure of a star forming region ever obtained by {\it Chandra}. However, these studies did not consider detailed dependencies of flare properties on stellar mass and had only a limited information on circumstellar disk indicators. Here, we update the COUP stellar mass values with the PARSEC-1.2S-based estimates, described in Paper~I. The $Spitzer$-based information on the presence/absence of disks is updated with the IRAC SED slopes ($\alpha_{IRAC}$) from the IRAC point source catalog of \citet{Megeath2012}. 

For each COUP flare, \citet{Getman08a} report a range of inferred loop heights based on the range of temperature-density slopes along the initial and late flare decay phases. To be compatible with the MYStIX/SFiNCs analyses (\S \ref{sec:flare_model}), the choice of flare durations and loop heights from \citet{Getman08a} is now limited to the flare initial decay stage only.

Table~\ref{tab:216_coup} lists the updated COUP star and super-flare properties used in this study. The format is similar to that of Table~\ref{tab:55_brightest} but excluding the rise-phase-based loop height ($L_{rise}$) and loop width ($\beta$) quantities since these quantities were not considered in Getman et al.

\section{PMS Super-Flare Results}  \label{sec:results2}

\subsection{Summary of MYStIX/SFiNCs Super-Flare Properties} \label{sec:summary_mystix_sfincs_flares}

While the atlas in Figure Set \ref{fig:atlas_example1} gives a detailed graphical view of the 55 bright MYStIX/SFiNCs, Table~\ref{tab:55_brightest} provides their scalar characteristics in a compact format.  They are again ordered by increasing $\beta$ values.  In some cases, loop heights and geometries could not be derived due to either insufficient flare decay/rise data or non-physical slopes on the temperature-density diagrams.  These are placed at the bottom of the table and atlas.  
The collective properties in Table~\ref{tab:55_brightest} and the atlas in Figure \ref{fig:atlas_example1} can be summarized as follows: 
\begin{description}
    \item [Host star mass and radius] The flaring stars have a wide range of  masses from 0.2~M$_\odot$ to  18~M$_\odot$.  Individual mass estimates may be quite uncertain, as they are estimated from dereddened infrared photometry calibrated to PARSEC-1.2S PMS evolutionary tracks \citep[][Paper I]{Bressan12,Chen14}. Here we are not  interested in exact masses for individual stars but in the trends of super-flare properties as functions of mass, where the mass scales are treated as flexible ``rubber bands'' dependent on the choice of methods and PMS evolutionary models. Radii are similarly estimated from the PMS evolutionary tracks.
 
    \item [Protoplanetary disk] The presence of a dusty disk around each star is inferred from the slope of the infrared spectral energy distribution.  Following Paper I, we classify stars with $\alpha_{IRAC} < -1.9$ as diskless stars and $\alpha_{IRAC} > -1.9$ as disk-bearing stars.  Close to half of the stars have disks and slightly more than half do not.  
    
    \item [Flare type] This classification indicates whether the {\it Chandra} exposure has caught the entire flare (`F') or only primarily its rise phase (`R') or decay phase (`D').  The flare peak properties in columns (8)-(11) may be underestimated for the 10 `D' flares listed at the bottom of the table.

   \item [Flare morphology]  The X-ray lightcurves are sometimes simple, with smooth exponential rise and decay phases (as in Figure~\ref{fig:atlas_example1}), and sometimes complex.  Complications include: double peaks; shoulders on a dominant peak; and a plateau after the rise phase.  Multiple peaks suggest violent reconnection in nearby magnetic loops, and a plateau  suggests continuing energy injection and chromospheric evaporation.
    
    \item [Rise and decay timescales] The flare rise times range from 0.4~ks to 40~ks with median $\sim 5$~ks.  Decay times range from 2.6~ks to 111~ks with median  $\sim 14$~ks.  In a handful of cases, the rise and decay timescales are similar but an asymmetrical fast-rise slow-decay shape is more common. In a quarter of the super-flares, the decay time is $\geq 3$ times the rise time.  
    
    \item [Peak powers] The peak luminosities range from $\sim~4\times10^{30}$ erg~s$^{-1}$ to $\geq 2.1 \times 10^{33}$ erg~s$^{-1}$ with median $7 \times 10^{31}$ erg~s$^{-1}$ in the $0.5-8$~keV {\it Chandra} band.  Peak emission measures range from $5 \times 10^{53}$ cm$^{-3}$ to $\geq 2 \times 10^{56}$ cm$^{-3}$ with median $6 \times 10^{54}$ cm$^{-3}$.  Recall these are likely lower limits for the 10 `D' class events listed at the bottom of Table~\ref{tab:55_brightest}.  The measures include the  highest super-flare values since the sample is chosen with a brightness criterion (Figure~\ref{fig:frd_flares}) but the shape of the distribution is not scientifically interesting as there is a wide range of distances to the MYStIX/SFiNCs star forming regions. 
    
    \item [Peak plasma temperatures] Plasma temperature at the peak of the emission measure ranges from 12~MK to $> 200$~MK with median 28~MK.  Temperatures in excess of $\sim 200$~MK are not well-constrained as the peak in the bremsstrahlung spectrum lies outside the range of the {\it Chandra} detection band \citep{Getman08a}.  Temperatures at the peak of the emission measure evolution are also provided in Table~\ref{tab:55_brightest} as they are needed in equation (\ref{eqn:loop_decay}). 
    
    \item[Loop heights and geometries] Loop heights estimated from decay brightness and temperature behavior range from 0.2~R$_\odot$ to 4~R$_\odot$ with median $1.4$~R$_\odot$. The loop heights are typically close to the stellar radius. Loop heights estimated from rise behavior are generally consistent with the heights estimated from decay flare phase. In cases of slow rise times (\S\ref{sec:variety}), inferred loop heights can reach $10-20$~R$_\odot$.  Some  MYStIX/SFiNCs super-flares have loop thickness-to-length ratios between 3\% and 10\% similar to solar flares, but half are considerably thicker with $\beta\sim 0.1-0.6$. Recall these values are derived with the assumption of a single loop with cylindrical geometry or multiple loops firing nearly simultaneously. High $\beta$ values may signal the presence of a flaring multi-loop structure. 
    
    \item[Time delays] The peaks of count rate, flare temperature, and emission measure are often not simultaneous. The most common occurrence is that the temperature peak precedes the emission measure peak by several hours. Similar but shorter time delays are commonly seen in solar and stellar flares \citep{Guedel1996, Reale2007, Getman2011}. They arise from rapid cooling of the plasma while the emission measure is still rising due to ongoing chromospheric evaporation in the coronal loop \citep{Benz2017}.
    
    \item[Plasma cooling processes] In some flares the radiative looses dominate the cooling process; in others the conduction dominates. In a few cases conduction dominates during the shorter rise and peak phases, but radiation cooling dominates during the longer decay phase.     
\end{description}

These results emerge from a uniform analysis and modeling of a large sample of energetic stellar super-flares. The MYStIX/SFiNCs stars with ages $<5$~Myr differ from main sequence stars and the Sun in important ways.  Most are rapidly rotating with fully convective interiors where magnetic dynamo processes must differ from solar-type tachoclinal dynamos. For example, in a broad analysis of dynamo processes of solar-type stars, \citet{Warnecke2020} conclude that an $\alpha^2$ mechanism combined with the R\"adler effect may be the main dynamo driver in rapidly rotating stars. The surface magnetic activity of such stars is orders-of-magnitude stronger than contemporary solar activity, evidenced here by the power of their X-ray flares. 

In the following subsections, we investigate several topics of particular interest relating to PMS super-flares:

(a) PMS super-flares show a wide variety of behaviors in their X-ray lightcurves and plasma models (\S\ref{sec:variety}). 

(b) Different plasma cooling processes are involved in these super-flares (\S\ref{sec:cooling_vs_plasma_temperature}). 

(c) In the context of Reale's solar loop model, a considerable fraction of PMS super-flares have thick rather than thin loop geometries, with $0.1 < \beta < 0.6$ rather than $\beta < 0.1$ as in solar magnetic loops (\S\ref{sec:thick_mass_dependence}). This phenomenon is correlated with host star mass.

(d) No evidence is found for flares arising in magnetic field loops connecting the PMS star with protoplanetary disks (\S \ref{sec:disk_dependence}). This supports conclusions based on super-flare energetics  reported in Paper I.

(e) PMS super-flares exhibit various scaling relationships reminiscent of those seen in solar flares (\S\ref{sec:solarlink}).

\subsection{Variety of Flare Evolutions} \label{sec:variety}

Roughly half of the X-ray lightcurves rise and fall smoothly.  Many of these are well-fit by exponential functions with faster rise phase and slower decay phase.  A few have symmetric lightcurves with similar rise and decay timescales. Median rise and decay timescales are $5$~ks and $14$~ks, respectively (Table~\ref{tab:55_brightest}).

No clear patterns are seen between flare brightness evolution --- rapid vs. slow timescales, smooth vs. jagged morphologies --- and other stellar or flare properties. This is reminiscent of the larger sample of COUP super-flares for which ``differences in flare morphology are not reflected in large differences in luminosity or temperature'' \citep{Getman08a}.  But two patterns unrelated to morphology are found and discussed below: a correlation between plasma cooling processes and temperatures (\S\ref{sec:cooling_vs_plasma_temperature}) and a correlation between the loop thickness and stellar masses (\S\ref{sec:thick_mass_dependence}). 

Nearly all well-modeled flares show the U-shaped evolution in the temperature-density diagram discussed for solar flares by \citet{Reale2014}.  The temperature increases quickly during the rapid rise phase, stays high while density increases around peak emission, followed by decreases in both temperature and density during the decay phase. But during flare \#12 the temperature keeps rising while emission measure starts decreasing; this leads to an inverted U-shape. Occasional flares have inverted U-shapes likely due to an incomplete decay coverage and/or contamination by another flaring event (e.g., flares \#30, 33).  

The most common discrepancy from the standard Reale model of flare evolution we see in some PMS super-flares are nonlinearities during the decay phase in the $T-\sqrt{EM}$ diagram. In such cases, our estimate of the slope $\zeta$ will be inaccurate or inapplicable to the more complex flare behavior of the super-flare event.

A number of theoretical and observational assumptions potentially affecting the $T-\sqrt{EM}$ trajectory may be too simplistic for PMS flares. On the theoretical side, such assumptions may include: consideration of a single flaring loop, invariant with time flaring loop geometry and volume, negligible energy and plasma flows during the flare decay phase, and spatially uniform but exponential in time loop heating distributions \citep{Reale1997}. On the observational side, such assumptions may include: single-temperature flaring plasma in a constant in time volume with spatially uniform density.

Around 50\% of the well-modeled PMS super-flares have plasma dominated by radiative cooling continuously during the flare and 30\% are dominated by conductive cooling continuously during the entire flare. The remaining 20\% typically show conductive cooling briefly around the peak emission and radiative cooling during the longer decay phase.  In solar flares, ``in the late phase, radiative cooling usually dominates, but considerable heat input is frequently observed'' \citep{Benz2017}.

Many of the MYStIX/SFiNCs super-flare X-ray lightcurves show secondary flares, most commonly during slow decay phases that have  timescales of $10-40$ ks.  Some secondary flares are relatively weak while others are comparable in intensity as the primary flare. In some cases, the secondary peak is more prominent in the temperature evolution plot than the X-ray luminosity plot.   Many COUP super-flares exhibit similar morphology, called ``step flares'' or ``double flares'' in \citet{Getman08a}.  \citet{Reale2004} observe and model a similar-morphology flare in the main sequence M5.5Ve star Proxima Centauri with age $\sim 5$~Gyr. They suggest that the secondary event may be occurred in an arcade adjacent to the primary flaring structure.  

A few of the smooth lightcurves show remarkably slow rise times.  For instance, super-flare \#20 labeled ``ic348 034435.37+321004.5\_1'' arising from a $\sim 3$~M$_{\odot}$ PMS star has $\tau_{rise}=36$ ks.  The single loop model fit indicates conductive cooling dominates during the entire event. Super-flare \#35 (flame 054134.34-020150.9) shows a slow rise with $\tau_{rise}$ around 40~ks with falling (rather than rising) temperature as the flux increases.  

These MYStIX/SFiNCs slow rise events recall PMS super-flares observed by other researchers. 
\begin{itemize}
\item[--] \citet{Grosso2004} observed a super-flare from an isolated diskless PMS star in the Orion clouds with {\it Chandra}. Here the X-ray emission rose for $\sim 20$ ks and followed by graduate increase in luminosity for $>10$ ks, all while remaining at a roughly constant temperature. The authors conclude that "The gradual phase ... is really unusual [and] may be explained by the rapid fall of the stellar magnetic field with height [where]  the reconnection site is continually moving up into a weaker field region".

\item[--] \citet{LopezSantiago2010} used {\it XMM-Newton} to analyze an X-ray flare with rise phase duration of 35~ks from an $\sim 8$~Myr old diskless star in the TW Hya Association. The temperature dropped slightly near the peak and inferred flaring loop height is $\sim 4$~R$_{\star}$.  The authors suggest the event "first involved mainly a single loop and then propagated to a loop arcade becoming a proper two-ribbon flare".   
\end{itemize}

A final mystery is why super-flares are seen from a handful of stars with masses around $5-15$~M$_\odot$.  These are generally O- and B-type stars that, in typical young clusters, are already on the main sequence with radiative interiors. X-ray flares from young OB-type stars have been reported in the past \citep{Stelzer05}. The origin of X-rays from late B stars is especially puzzling since they possess neither the strong winds producing X-rays from early B- and O-type stars nor the convective envelopes needed for the X-ray production through magnetic dynamos in lower-mass stars.  Compared to the surface magnetic fields of PMS stars which reach up to a several kilo-Gauss \citep{Sokal2020},  the fields in young Herbig Ae/Be stars are thought to be much weaker \citep[$<100-200$~G,][]{Hubrig2015}. 

However, this should not be definitively viewed as a puzzle.  It is quite possible that the super-flares arise from lower mass PMS companions of these more massive stars.  This is a persuasive explanation for the characteristic X-ray emission from some B stars \citep{Schmitt1985,Stelzer05, Stelzer09}.

\begin{figure} [hb]
\epsscale{0.95}
\plotone{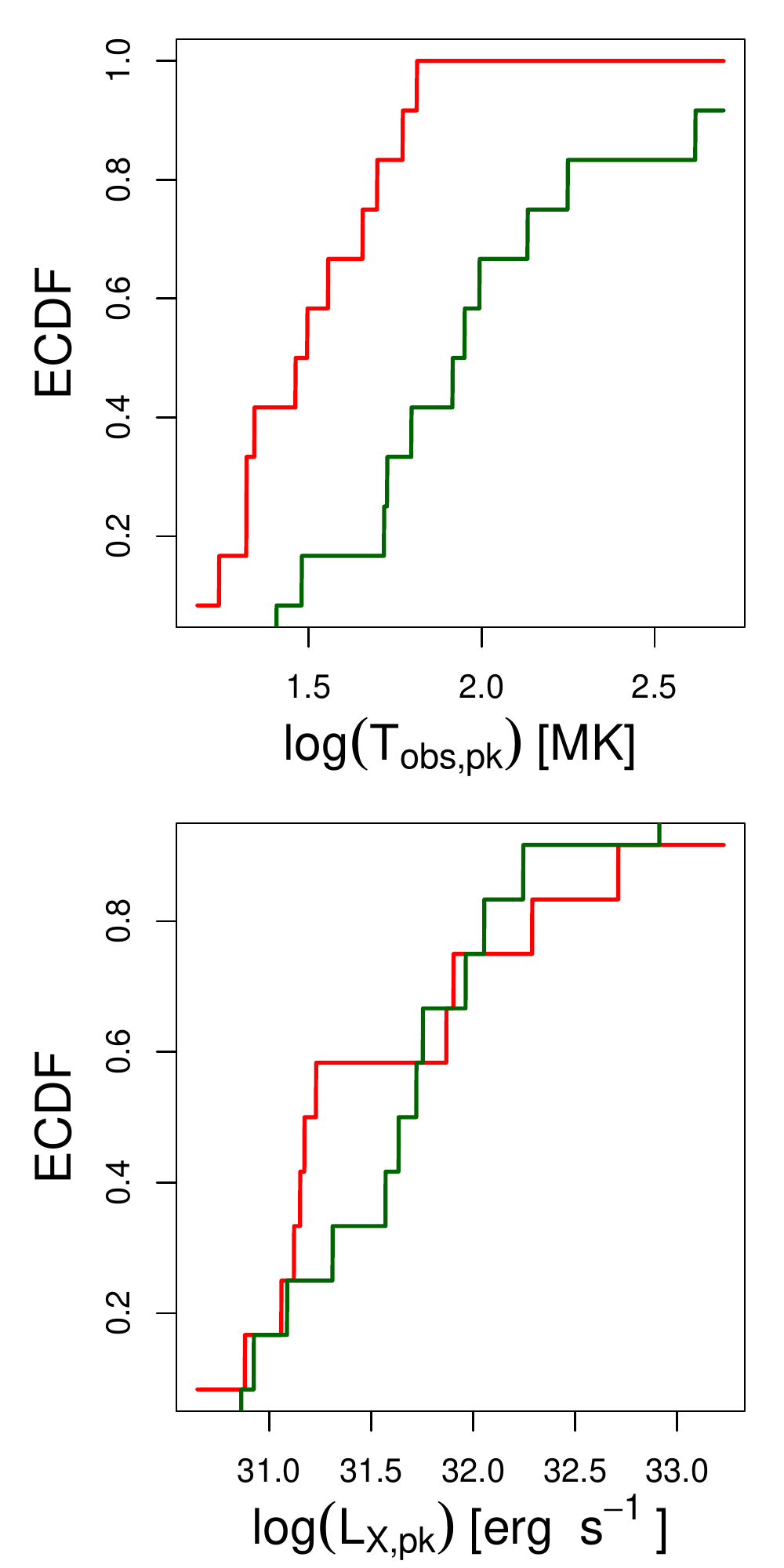}
\caption{Empirical cumulative distribution functions of flare peak plasma temperature and X-ray luminosity stratified by the dominant cooling process. Super-flares dominated by radiative losses are shown in red and super-flares dominated by conductive losses are shown in  green.  \label{fig:bright_superflares_losses_strata}}
\end{figure}

\subsection{Cooling processes and plasma temperature} \label{sec:cooling_vs_plasma_temperature}

Out of the 24 MYStIX/SFiNCs super-flares with available temporal profiles of the radiation-conduction cooling timescales, 12 have flare evolution entirely dominated by the radiative cooling, 7 are entirely dominated by conductive cooling; and 5 exhibit the switch from conduction during the rise phase to radiation during the decay phase.  

According to the equations (\ref{eq:cooling_timescales})-(\ref{eq:radiation_timescales}),  conduction governs when the plasma temperatures are high while radiation is important when plasma temperatures are low and/or X-ray luminosity is high \citep{Cargill1995,Benz2017}. This relationship is validated in Figure~\ref{fig:bright_superflares_losses_strata} showing our modeled peak plasma temperature and X-ray luminosity for two sub-samples of these super-flares when the losses are governed by radiation or, at least partially, by conduction. The plasma temperatures are clearly  higher in super-flares dominated by the conductive losses ($P \simeq 1$\% with the two-sample Anderson-Darling test).

Since our bright MYStIX/SFiNCs super-flares are ones among the most X-ray energetic and luminous PMS super-flares known (Figure \ref{fig:frd_flares}), we would expect the radiative losses to dominate. However, even in the most luminous X-ray flaring events, the conduction may play a key role when plasma temperature exceeds roughly 50 MK.

We find consistent evidence that flare plasma cooling loss mechanism is related to the mass of the host star in the sense that conductive cooling may be especially important in super-flares produced by more massive PMS stars. Flare loop widths $\beta$ positively correlate with host star masses (\S\ref{sec:thick_mass_dependence}). Examination of the atlas shows that super-flares ruled by conductive losses are also preferentially found in flares associated with high $\beta$ values.  Flare temperatures may also be hotter in high mass stars (\S \ref{sec:coup2}).

\subsection{Thick Loops and Stellar Mass} \label{sec:thick_mass_dependence}

\begin{figure} [hb]
\epsscale{0.95}
\plotone{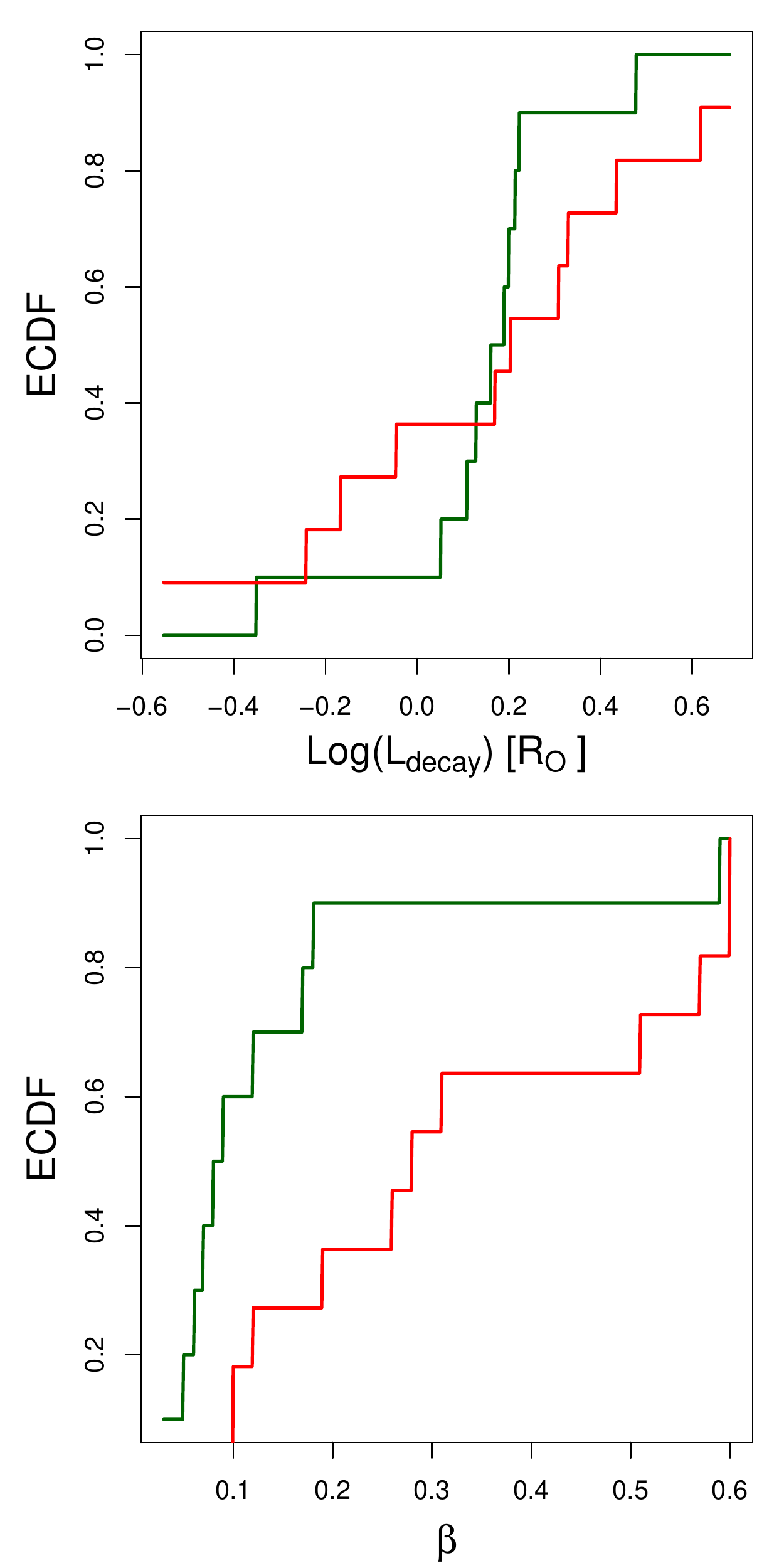}
\caption{Loop geometries for the bright super-flares.  Empirical cumulative distribution functions of loop height and thickness stratified by stellar mass; super-flare samples with $M<1$~M$_{\odot}$ (10 flares) and $M>1$~M$_{\odot}$ (11 flares) are color-coded in green and red, respectively.  \label{fig:bright_superflares_mass_strata}}
\end{figure}

Perhaps the most unexpected result from our analysis of MYStIX/SFiNCs super-flares concerns the measure $\beta$ of loop thickness, the ratio of loop's cross-sectional radius to height, in the flaring loop model of \citet{Reale2007}.  The results are presented in the flare atlas and Table~\ref{tab:55_brightest} and visualized in the bottom panel of Figure \ref{fig:bright_superflares_mass_strata}.  First, only one-fourth of the modeled super-flares have $\beta < 0.1$, values characteristic of solar flares \citep{Handy1999,Aschwanden2000}. Other inferred values of $\beta$ range from 0.1 to nearly 0.7.  Second, super-flares arising in young stars with masses exceeding 1~M$_\odot$ have thicker coronal loops.  The statistical significance is $P \simeq 1$\% with the two-sample Anderson-Darling test.  The host star's mass affects loop thickness but not loop height (Figure \ref{fig:bright_superflares_mass_strata}, top panel).  This result implies higher flaring volumes and likely larger loop footprint areas on the stellar surfaces of higher-mass stars.

It is possible that high-$\beta$ solutions to the model of \citet{Reale2007} may indicate that multi-loop arcades, rather than single large loops, are involved. Combined with the host mass effect, this would suggest that more massive PMS stars harbor flaring coronal arcades arising from  larger active regions with strong magnetic flux.  

\begin{figure*}[ht]
\epsscale{1.15}
\plotone{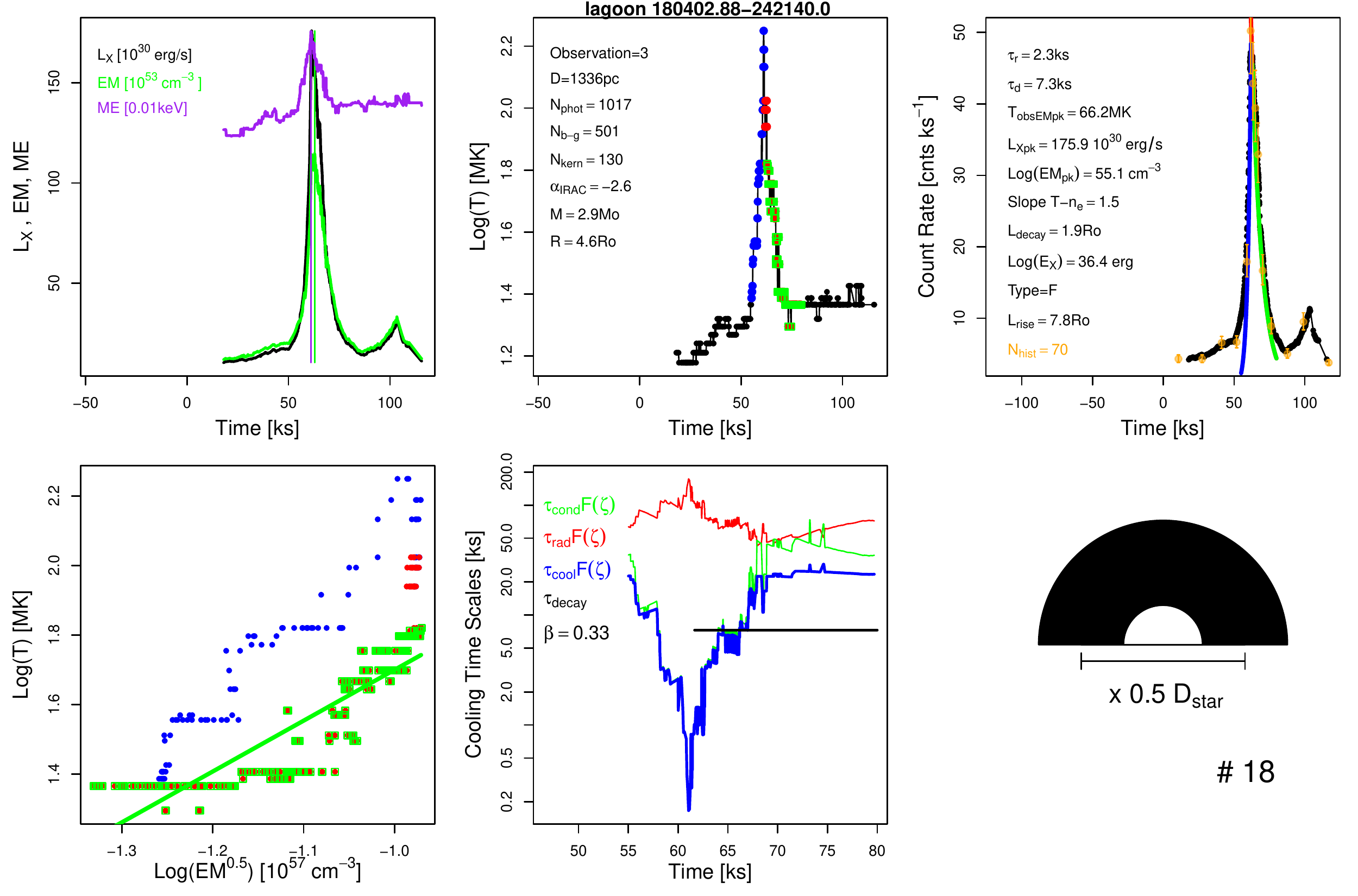}
\caption{Modeling for a super-flare from a 3~M$_{\odot}$ host star with an inferred thick loop geometry with high $\beta$ value.  See Figure~\ref{fig:atlas_example2} and \S\ref{sec:masme_method} for a detailed description.  \label{fig:atlas_example2}}
\end{figure*}

Figure~\ref{fig:atlas_example2} provides a detailed view of the flare evolution and inferred loop geometry for a high-mass, high-$\beta$ super-flare. Here the host PMS star in the Lagoon Nebula region has mass $\sim 3$~M$_\odot$. The peak X-ray luminosity and emission measure are $\sim 10$ times higher compared to the 1~M$_{\odot}$ star in Figure~\ref{fig:atlas_example1}; this is a mega-flare.  The peak temperature is somewhat hotter than in the less luminous flare, although the flare rise and decay timescales are similar.  The resulting loop model requires a high volume represented by $\beta = 0.33$.  Cooling is dominated by conduction rather than radiation throughout the decay phase of the flare.  

In this small MYStIX/SFiNCs sample with available loop geometries and stellar masses (21 flares), we do not see statistically significant trends between mass strata and other flare properties.  However when the remaining MYStIX/SFiNCs super-flares and super-flares from the Orion Nebula Cluster are added (\S \ref{sec:coup2}, panels c$-$g in Figure~\ref{fig:compare_with_coup}), clear correlations are seen between stellar mass and flare peak emission measure and total energy; and a correlation with peak plasma temperature may possibly be present. No correlation between mass and loop height is seen.

These results suggest that the flaring geometry and power of intermediate- and high-mass PMS stars may differ from lower-mass PMS stars.

\subsection{No Relation to Protoplanetary Disks} \label{sec:disk_dependence}

The slope of the infrared spectral energy distribution is widely used as a photometric measure of dusty protoplanetary disks.  Following Paper I, we classify PMS stars with $\alpha_{IRAC} > -1.9$ as `disk-bearing' and those with steeper slopes as 'diskless' (Class III stars).  This criterion is similar to the classification in \citet{Lada2006} and \citet{Hernandez2007} where our disk-bearing stars are accreting Class I and II stars, and our diskless stars include both Class III and the rare class of `anemic' disks. For instance, Figure 9 of \citet{Lada2006} shows that 68\% of known actively accreting objects in IC~348 have $\alpha_{IRAC} > -1.8$. But 11\% of known accretors are found to be associated with `anemic' disks ($\alpha_{IRAC} < -1.8$). A small fraction of MYStIX/SFiNCs/COUP flare-host stars with $\alpha_{IRAC} < -1.9$ may still posses anemic disks and accretion.

A number of observational studies have suggested that the model inferences of enormous loop heights is evidence that the flaring magnetic structures connect the star to the inner disk.  In an early COUP study,  \citet{Favata2005} found that loops associated with some of the most powerful COUP flares reach heights up to several stellar radii, and suggested that such long loops may have difficulty withstanding centrifugal forces unless anchoring with one footpoint at the stellar surface and the other at the inner circumstellar disk.  Centrifugal stripping of large coronal loops in rapidly rotating magnetically active stars ha bseen studied in detail, including for PMS stars \citep{Unruh1997}. The observed locus of stellar X-ray luminosity versus rotation period for PMS members of the 13~Myr h~Per cluster is found to be consistent with the predictions of the centrifugal stripping model \citep{Argiroffi2016}. Other observational X-ray studies find long flaring coronal loops for disk-bearing stars \citep{Giardino2006, Giardino2007, McCleary2011, Lopez-Santiago2016, Reale2018}.

However, in a larger sample of COUP super-flares (\S \ref{sec:COUP_superflares_description}, \ref{sec:coup2}), \citet{Getman08a,Getman08b} found that coronal loops with heights up to several stellar radii are indistinguishably present in both disk-bearing and diskless stars. They report tentative evidence that loop heights never exceed the inner disk radius in disk-bearing stars.  This does not necessarily imply star-disk magnetic connection, as it could arise from disk confinement of T Tauri magnetospheres with solar-like loops.  These results are consistent with the predictions of X-ray coronal extent in the presence of PMS disks reported by \citet{Jardine2006} who combined empirical X-ray emission measures with 3-dimensional models of T Tauri magnetospheres based on Zeeman mapping of surface magnetic fields.

In the present study, we confirm the absence of a relationship between loop model parameters and protoplanetary disks for the 55 bright MYStIX/SFiNCs super-flares. The properties of the bright MYStIX/SFiNCs super-flares, such as loop geometry, flare duration, flare peak emission measure and plasma temperature, and flare energy do not depend on the presence/absence of disks (Table \ref{tab:55_brightest}; figure is not shown). However, we stress that the short exposure times of the {\it Chandra} observations of MYStIX/SFiNCs (compared to COUP) reduce the possibility of detecting extremely long events, and hence very long flaring loops (\S \ref{sec:coup2}).

\begin{figure*}
\epsscale{1.15}
\plotone{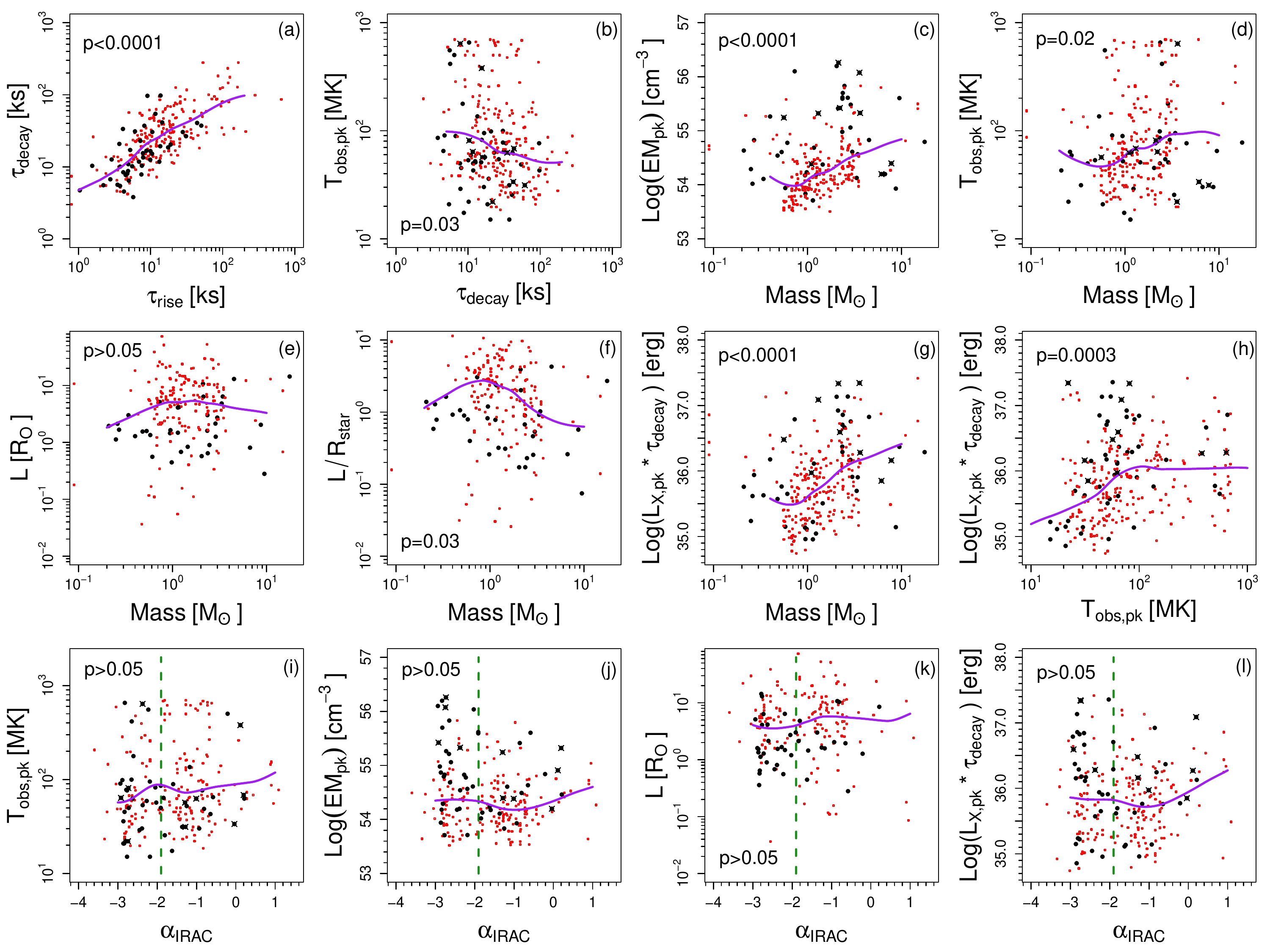}
\caption{Bivariate scatterplots of several properties of the 55 bright MYStIX/SFiNCs flares (black points) with comparison to the COUP X-ray flares  \citep[][red points]{Getman08a,Getman08b}. Black $\times$ symbols represent lower limits for flares that were not fully captured in the MYStIX/SFiNCs exposures. Local regression fits \citep{Loader20} are shown as magenta curves. Dashed vertical lines in the bottom panels denote the chosen boundary between disk-bearing (right) and diskless (left) stars.  Legends give p-values from Kendall's $\tau$ bivariate nonparametric correlation test.  Panel descriptions: (a) Flare rise versus decay timescales. (b) Observed peak flare temperature versus decay timescale. (c,d,e,f) Peak flare emission measure, peak flare temperature, flare loop height, flare loop height normalized by stellar radius as functions of stellar mass. (g,h) Flare energy as a function of stellar mass and peak flare temperature. (i,j,k,l) Peak flare temperature, peak flare emission measure, flare loop height, and flare energy as functions of infrared spectral energy distribution slope $\alpha_{IRAC}$.  
 \label{fig:compare_with_coup}}

\end{figure*}

No disk relationships emerge when these bright MYStIX/SFiNCs super-flares are combined with the COUP super-flares. The bottom row of  Figure~\ref{fig:compare_with_coup} shows no correlation between peak plasma temperature, peak emission measure, loop height, or X-ray energy and the $\alpha_{IRAC}$ infrared slopes.  Statistical tests using Kendall's $\tau$ show no correlations. If we stratify the sample into disk-bearing and diskless stars at the $\alpha_{IRAC} = -1.9$ boundary, Anderson-Darling two-sample tests again show no difference (bottom row of Figure~ \ref{fig:compare_with_coup}).  This is despite the fact that the MYStIX/SFiNCs flare emission measures and energies are systematically higher, and loop heights are lower, than those of the COUP super-flares, which is explained in \S \ref{sec:coup2} as a data selection effect.

\begin{figure*}[ht!]
\epsscale{1.1}
\plotone{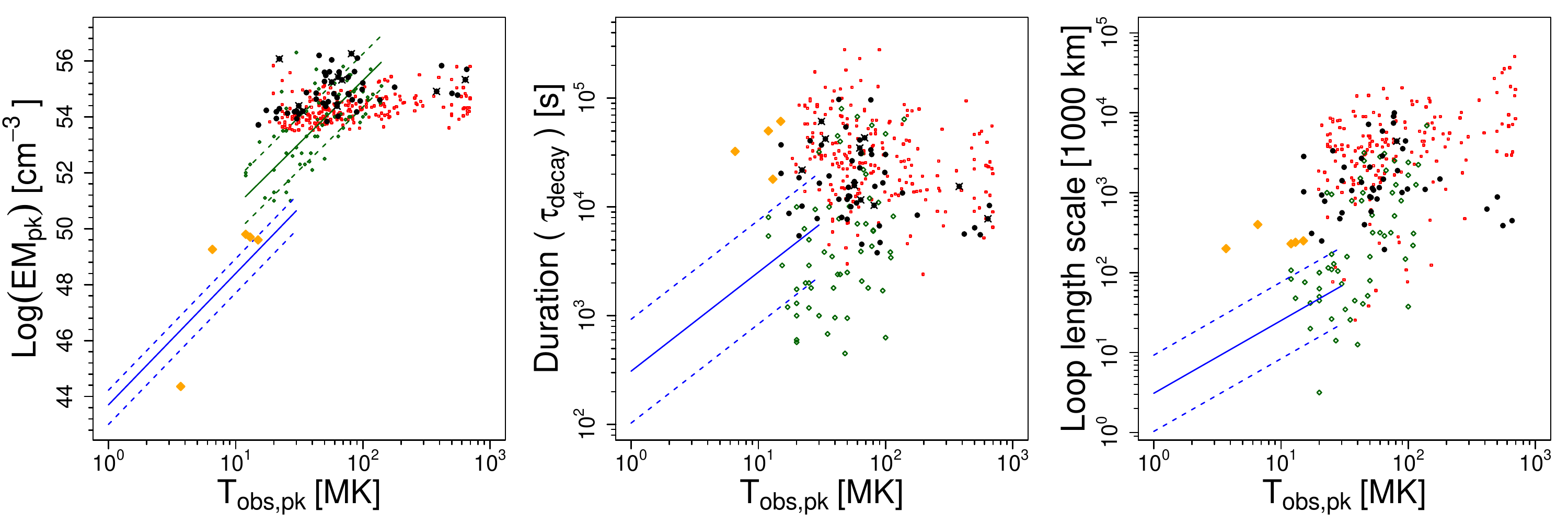}
\caption{Peak flare emission measure (left), flare decay timescale (middle), and loop height (right) as functions of peak flare temperature. The 55 bright MYStIX/SFiNCs super-flares (black points) with lower limits on $EM_{pk}$ and $T_{obs,pk}$ for partially captured flares (black $\times$s). COUP flares are marked by red points and stellar flares (both from young and older stars) compiled by \citet{Gudel2004} are marked by green points with their linear regression fit (in the left panel) shown as green lines. Regression fits with 1-$\sigma$ range  for solar flares compiled by \citet{Aschwanden2008} are shown as blue lines.  Solar LDEs compiled by \citet{Getman08a} are shown as orange diamonds. \label{fig:compare_with_olderstars}}
\end{figure*}

\section{Comparison to Other Stellar Flares} \label{sec:comparison_among_flares}

\subsection{Orion Nebula Cluster X-ray Super-Flares} \label{sec:coup2}

Figure \ref{fig:compare_with_coup} compares the COUP flares (red points) to our sample of 55 bright MYStIX/SFiNCs super-flares (black points) in a variety of bivariate scatterplots.  Recall that the MYStIX/SFiNCs sample includes partially captured flares with likely lower limits on $T_{obs,pk}$, $EM_{pk}$, and $L_{X,pk}$; these values are shown with black $\times$ symbols. We preferentially plot $L_{decay}$ over $L_{rise}$ loop heights for the MYStIX/SFiNCs sample.  Since \citet{Getman08a} did not derive coronal widths for the COUP flares but rather assumed a fixed value $\beta = 0.1$, comparison involving  $\beta$ values are omitted.

Some scientifically uninteresting selection effects are present in the comparison between MYStIX/SFiNCs and COUP super-flares.  First, the bright MYStIX/SFiNCs super-flares lack extremely long events seen in the COUP sample (panels a and b) due to the shorter MYStIX/SFiNCs {\it Chandra} exposure times. The distribution of COUP rise and decay timescales is more realistic than the MYStIX/SFiNCs distributions even with statistical correction for truncation with the Kaplan-Meier estimator. Second, due to the  selection criterion $>$1000-photon flare-host {\it Chandra} observation in the MYStIX/SFiNCs sample, fewer heavily absorbed super-flares with very hard spectra are included.  This leads to fewer `superhot' flares \citep[discussed in][] {Getman08a} compared to the COUP sample. Third, based on equation (\ref{eqn:loop_decay}), the shorter decay timescales and lower temperatures lead to a bias that MYStIX/SFiNCs sample has fewer super-flares with very large loop heights ($L \simeq 2-10$~R$_{star}$) compared to the COUP sample.  This is seen in panels e and f.  

However, some selection effects are helpful.  Some of the 55 bright MYStIX/SFiNCs super-flares studied are substantially more powerful than most of the COUP super-flares, as seen in panels c, g, h, j, and l of Figure~\ref{fig:compare_with_coup}. This arises from the greater \'etendue of the MYStIX/SFiNCs surveys. The areas covered by $Chandra$ in $\sim 40$ star forming regions include $\sim 112,000$ PMS stars with masses above 0.1~M$_\odot$ (Table 1 in Paper I).  With typical $Chandra$ exposures around 74~ks, the MYStIX/SFiNCs survey detects $\sim 5$ times more super-flares than the COUP observation of the ONC (1,086 $vs.$ 216).  The limit here to events with $>1000$-photon $Chandra$ observations restricts consideration to some of the most luminous super-flares. 

The joint MYStIX/SFiNCs-COUP sample in Figure~\ref{fig:compare_with_coup} shows several clear correlations between modeled properties.  The peak flare emission measure and total flare energy scales with stellar mass (panels c and g), and there is a tentative correlation between the peak flare temperature and stellar mass (panel d).  These are not expected effects; they are discussed below in \S\ref{sec:disucssion_mass_dependence}. 

We also see a correlation between total flare energy and peak plasma temperature (panel h).  It is well-known that at lower energies more powerful solar flares have hotter plasma \citep{Aschwanden2008}, and the effect is seen in the full sample of COUP stars at intermediate energies \citep[see section 7 of][]{Preibisch05}.  Finally, we have discussed in \S\ref{sec:disk_dependence} that flare properties do not correlate with the presence/absence of disks (panels i, j, k,and l).

\subsection{Comparison with X-ray Flares from Older Stars} \label{sec:older_stars2}

Figure~\ref{fig:compare_with_olderstars} compares single-loop model characteristics of PMS super-flares, shown in black for MYStIX/SFiNCs and in red for COUP as in Figure~\ref{fig:compare_with_coup}, with flares from older stars. Stellar flares shown in green, given for young stars and older active stars, are from the compilation of the most powerful and hottest stellar flares assembled by \citet{Gudel2004}.  The loci of typical solar flares are shown as the blue lines \citep{Aschwanden2008}. Similar plots for the COUP PMS super-flares alone are shown by \citet{Getman08a} with similar discussion.

The strongest difference between PMS and main sequence flaring in Figure~\ref{fig:compare_with_olderstars} is that the plasma in PMS super-flares are typically $2$ to $>$20 times hotter than seen in the most luminous solar flares, and can be hotter than main sequence stellar flares.  The presence of `superhot', $T_{obs,pk}>100$~MK, PMS flares in the COUP observation was noted by \citet{Favata2005} and \citet{Getman08a}.  

The PMS super-flares do exhibit the statistical association between plasma emission measure and temperature at the peak of the flare ($P \ll 1$\% with Kendall's $\tau$ test) that is seen in both solar and older star flare samples. But the PMS super-flare power law slope $\alpha = 0.4$ in the $EM_{pk} \propto T_{obs,pk}^{\alpha}$ relation is much shallower than that for the solar-stellar flares, $\alpha = 4.7$ \citep{Aschwanden2008} that is close to the theoretical prediction of $\alpha = 4.3$ based on the solar loop scaling laws \citep[][RTV]{Rosner1978}. 

However, we stress that PMS X-ray flares with lower temperatures and emission measures are likely present on PMS stars, but are either unresolved or too faint for analysis with the \citet{Reale2007} loop model. Paper~I establishes the flare energy distribution of $dN/dE_{flare} \propto E_{flare}^{-2}$ for the PMS super-flares. If weaker flares follow a similar distribution, numerous such flares are expected to blur together, contributing to the continuous ``characteristic'' emission commonly seen in PMS stars \citep{Wolk05}.  Therefore, the intrinsic slope in the $EM_{pk} \propto T_{obs,pk}^{\alpha}$ relation for PMS flares may be steeper than $\alpha=0.4$ but can not be measured here based on the modeled bright super-flares. 

The decay phase timescales of PMS super-flares, typically between 3 and 30 hours, are longer than seen on most solar and stellar flares (Figure~\ref{fig:compare_with_olderstars}, panel b). But a few solar Long-Duration Events (LDE, orange points) and older star flares have similarly long durations.  Solar LDEs are relatively rare with exceptionally long-lasting X-ray arches and streamers, though modest emission measures. LDEs are the largest known associated flaring solar structures with heights reaching up to 50\% of the solar radius \citep{Shibata2011}. 

A few extreme flares from older stars have loop heights comparable to PMS super-flares, but flare lengths exceeding a solar radius have never been seen on the Sun, even from LDEs (panel c). The modeled PMS super-flare loop heights show a correlation with peak plasma temperature according to the equation \ref{eqn:loop_decay}.  This in turn is a consequence of the solar RTV loop scaling laws coupled with the relationship of the flare decay timescale with loop's volumetric heating \citep{Serio1991} and possible presence of prolonged heating in the loop \citep{Reale1997}.

\subsection{Solar Flares and Magnetic Flux Relations} 
\label{sec:solarlink}

\begin{figure} [ht]
\epsscale{1.15}
\plotone{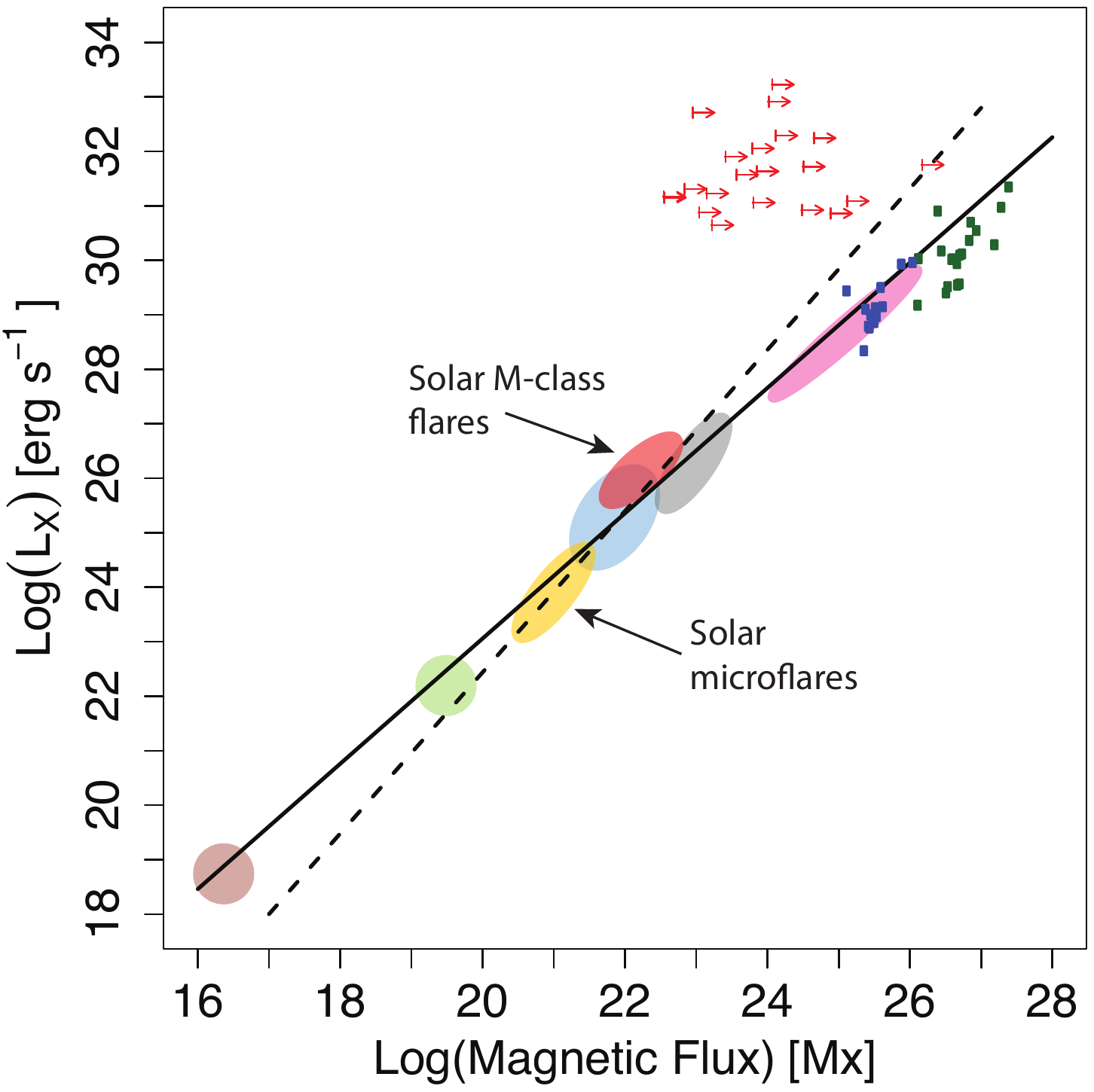}
\caption{X-ray luminosity as a function of magnetic flux. Lower limits on the magnetic flux for our bright MYStIX/SFiNCs super-flares with available information on loop geometries and stellar radii are indicated by the red arrows.  The solid and dashed black lines are the relations $L_X \propto \Phi^{m}$ with $m=1.15$ \citep{Pevtsov2003} and $m=1.48$ \citep{Kirichenko2017}, respectively. These lines fit various magnetic elements emitting X-rays.  From \citet{Pevtsov2003}, these include the quiet Sun (brown), solar X-ray bright points (light green),  solar active regions (cyan), solar disk averages (grey), and old G, K, M dwarfs (magenta). Solar microflares (golden oval) are from \citet{Kirichenko2017} and solar M-class flares (red oval) are from \citet{Su2007}. Individual solar-type dwarfs (blue points) are from \citet{Kochukhov2020}. Individual Orion and Taurus PMS stars (dark green points) are from \citet{Sokal2020}. \label{fig:pevtsov_plot}}
\end{figure}

PMS super-flares and solar flares span very different energy ranges, $E_{PMS} = [10^{34} - 10^{38}]$~erg (Paper~I) and $E_{solar} = [<10^{24} - 10^{32}]$~erg \citep[e.g.,][]{Okamoto2021}. Despite this enormous difference, one can draw many common, often flare-model-independent features pertaining to both types of flares:
\begin{description}

    \item[Loop Anchoring At Stellar Surface] In \S \ref{sec:disk_dependence} and Paper I, we report no observational evidence for star-disk flaring PMS coronal structures.  This suggests that, as on the Sun, PMS flaring loops have both foot-points rooted in the stellar surface.  

    \item [Flare Evolution] The observed X-ray (and other bands) light-curves for both types of flares often exhibit common patterns, such as exponential fast-rise and slow-decay. Time delays are common, with the temperature peaks often preceding the emission measure peaks (\S \ref{sec:summary_mystix_sfincs_flares}).
    
    \item [Energy Distribution] The solar flares and PMS super-flares follow the similar energy distribution law $dN/dE \propto E^{-2}$ over $>$14 orders of magnitude energy range (Paper I).
    
    \item [Heating and Cooling Processes] Nearly all PMS super-flares analyzed here show the U-shaped evolution in the temperature-density diagram (\S \ref{sec:summary_mystix_sfincs_flares}) characteristic of solar flares and explained by the single-loop plasma model of \citet{Reale2007}.
    
    \item [Analogs of Solar Long Duration Events (LDEs)] LDEs are outliers from the scaling laws of the typical solar flares (Figure \ref{fig:compare_with_olderstars}). Among solar flares, the LDE properties are closest to those of the PMS super-flares. \citet{Getman08a} propose that PMS super-flares are enhanced analogs of LDEs.
    
    \item [Scaling Relationships] Inferred positive correlations for a number of PMS super-flare relationships, such as $\beta - M$ (the only flare-model-dependent relation), $EM_{pk} - M$, $T_{obs,pk} - M$, $E_X - M$, $E_X - T_{obs,pk}$,  $EM_{pk} - T_{obs,pk}$ (\S \ref{sec:thick_mass_dependence}, \ref{sec:coup2}, \ref{sec:older_stars2}), suggest that more energetic and X-ray luminous PMS super-flares are associated with hotter plasma and larger loop foot-print areas. Qualitatively similar conclusions can be reached for the more luminous solar flares based on the solar flare scaling laws depicted in Figure \ref{fig:compare_with_olderstars} and Figure \ref{fig:pevtsov_plot} discussed below. The measured power law slopes may differ between the solar and PMS flares, but the incompleteness of data on weaker PMS flares may play an important role here (e.g., the $EM_{pk} - T_{obs,pk}$ correlation discussed in \S \ref{sec:older_stars2}).  
\end{description}

However, PMS super-flares appear to violate one important scaling law.  \citet{Pevtsov2003} report a universal, approximately power law,  relationship between the X-ray luminosity ($L_X$) and magnetic flux,
\begin{equation} \label{eqn:Lx_magflux}
    L_X \propto \Phi^m ~~ {\rm with} ~~ m=1.15,
\end{equation}
where $\Phi \sim B \times A$ and $A$ is the surface area through which the magnetic field with strength $B$ passes (Figure~\ref{fig:pevtsov_plot}, solid black line).  The relationship seems to apply across an extraordinary range of $\sim 13$ orders of magnitude in X-ray luminosity.  The relationship connects low-level X-ray emission and magnetic fluxes from individually resolved small-scale solar elements, mid-level emission from solar active regions, and high-level emission from main sequence and characteristic (not super-flaring) emission from PMS stars. Solar microflares are reported to have a steeper relation (dashed black line) with the slope of $m=1.48$ \citep{Kirichenko2017}. More powerful solar M-class flares appear to follow the $F_X \propto \Phi^{1.1}$ trend, where $F_X$ refers to a $GOES$ flux in the soft X-ray band \citep{Su2007}.

To this plot, we add 14 individual solar-type stars using their magnetic field strengths from \citet{Kochukhov2020} and X-ray luminosities and stellar radii from \citet{Vidotto2014}. We also add 19 Taurus and Orion PMS stars with characteristic X-ray luminosities in \citet{Gudel2007b} and \citet{Getman05}, using magnetic field strengths from \citet{Sokal2020}. These stars indeed seem to follow the nearly linear relationship from \citet{Pevtsov2003}\footnote{Small systematic shifts in $\log(L_X)$ among various magnetic elements may be expected due to the choice of slightly different energy bands: $[0.3-4.4]$~keV for solar elements and $[0.1-2.4]$~keV for some G, K, M stars in \citet{Pevtsov2003}; $[0.3-4.4]$~keV for solar microflares in \citet{Kirichenko2017}; $[1.5-12]$~keV for solar M-class flares in \citet{Su2007}; $[0.3-10]$~keV for Taurus PMS stars \citep{Gudel2007b}; and $[0.5-8.0]$~keV for the MYStIX/SFiNCs super-flares and Orion PMS stars \citep{Getman05}.}. 

There is debate on the reliability of the relationship in equation (\ref{eqn:Lx_magflux}).  \citet{Zhuleku2020} summarize observational studies that report linear or non-linear dependence of $L_X \propto \Phi^{m}$ with slopes ranging from 0.89 to 2.68\footnote{Notice that the slope $m=2.68$ quoted by \citet{Zhuleku2020} was taken from the study of solar-type stars by \citet{Kochukhov2020}, who in fact report such a slope value with regards to a different relationship, $L_X/L_{bol} \propto B^{2.68}$.}. They then develop an analytical formulation to explain the $L_X \propto \Phi^m$ relationship by combining the Rosner-Tucker-Vaiana scaling laws \citep{Rosner1978,Serio1981} with theories on coronal heating \citep{Fisher1998} and X-ray radiation from optically thin thermal plasma \citep{Mewe1985}.

Within the framework of this analytical model, \citet{Zhuleku2021} perform three-dimensional MHD simulations of an active region by mimicking the coronal heating through ``the Ohmic dissipation of currents induced by photospheric magneto-convective motions'' (nanoflare heating). The size of the active region was kept constant, but the surface magnetic field strength was varied from 1~kG to 20~kG. The simulations with high magnetic field strengths specifically oriented to magnetically active stars predict a steep $L_X \propto \Phi^{m}$ relationship with $m=3.44$. 

We can place the MYStIX/SFiNCs PMS super-flares into this picture in a semi-quantitative fashion.  From the data listed in Table~\ref{tab:55_brightest}, the minimum magnetic field strength required to confine the plasma assuming magneto-thermal pressure equipartition and the simple geometry of the \citet{Reale2007} loop model is approximately 
\begin{equation}
B_{eqp} = \sqrt{8 \pi k_b n T}  
\end{equation}
where plasma density is $n=(EM/V)^{0.5}$ and flaring volume is $V=2 \pi \beta^2 L^3$.  If we assume a dipolar large-scale magnetic topology, then $B(L) \simeq B_{ph}/(L/R_{\star} + 1)^3$ and a lower limit on the surface magnetic field strength can be estimated.  The magnetic flux producing the loop is $\Phi = B_{ph} \times Area$ where the surface area covered by the loop is $Area = \pi \beta^2 L^2$.  

Applying the properties found for the MYStIX/SFiNCs super-flares in Table~\ref{tab:55_brightest} based on the \citet{Reale1997} loop model, we find median loop height $L/R_{\star} \sim 0.8$, typical loop's cross-sectional area $10^{21}$~cm$^2$, equilibrium loop magnetic field $B_{eqp} \sim 200$~G, and surface magnetic field $B_{ph}\sim 1$~kG. This minimum surface field strength is $\sim 2$ times lower than the typical field strengths, averaged across entire stellar surfaces of PMS stars, derived from Zeeman broadening methods \citep{Yang2011, Sokal2020}. 

It is possible that we further underestimate the surface magnetic field strengths for some higher-mass PMS stars by assuming the large-scale dipolar topology. For example, \citet{Gregory2014,Gregory16} suggest that in higher-mass PMS stars that start developing radiative cores their magnetic topologies become more complex, changing from dipolar to octopolar. The assumption of octopolar geometry ($B(L) \simeq B_{ph}/(L/R_{\star} + 1)^5$) would further increase our surface field estimates by a factor of 4.

The inferred minimum magnetic fluxes of our PMS super-flares then range between $10^{23} - 10^{25}$~Mx.  These are plotted against our super-flare peak X-ray luminosities $L_{X,pk}$ as red arrows in Figure~\ref{fig:pevtsov_plot}.  The loops can lie to the right of these locations if they are out of equilibrium in the sense that the plasma is overconfined by the magnetic field. 

We thus find in Figure~\ref{fig:pevtsov_plot} that the MYStIX/SFiNCs PMS super-flares simultaneously have X-ray luminosities $3-4$ orders of magnitude above, and minimum magnetic fluxes $2-3$ orders of magnitude below,  the predictions of the shallow $m=1.15$ relationship reported by  \citet{Pevtsov2003}. The shift of the PMS super-flares to the line of the solar microflares from \citet{Kirichenko2017} requires surface magnetic field strengths of $B_{ph} = 100$~kG in PMS star spots responsible for such energetic super-flares, a huge $25-100$-fold increase above the typical average PMS surface magnetic field levels. While this may seem unrealistic, one may recall that on the Sun itself the magnetic field strengths of a few kG in solar spots are by factors of 10-20 higher than the average surface field strength of 200~G \citep{Kochukhov2020}. 

Overall, our results favor a steeper $L_X \propto \Phi^m$ relationship than seen in the Sun and other stars.  We thus empirically support the calculation of steeper $m$ values calculated by \citet{Zhuleku2021}. Their simulations specifically address strong magnetic flaring associated with the magnetic field strength in a starspot of up to $B_{ph} = 20$~kG.
   
\section{Discussion} \label{sec:discussion2}
We study here some of the most powerful X-ray flares ever recorded and modelled from any stellar population. The current work complements the previous COUP studies of \citet{Favata2005} and \citet{Getman08a} by providing detailed modeling of some of the most powerful PMS flares originating outside the Orion Nebula Cluster. We discuss here three aspects of the results that address astrophysical issues. 

\subsection{Applicability of the Solar Single Loop Model}
\label{sec:solar_applicability}

Our modeling of the {\it Chandra} X-ray data, like many previous studies, relies on assumptions that PMS flaring shares the same the fundamental physical processes and geometries as solar flares.  We find observational and theoretical evidence supporting this similarity, but also indications of significant differences. 

Observationally, PMS super-flares with their long coronal loops show some common properties to rare long-duration eruptive solar flare that are accompanied by large-scale coronal mass ejections.  The solar events exhibit X-ray emitting arches and streamers with altitudes reaching up to a few hundred thousand kilometers (Figure~\ref{fig:compare_with_olderstars} c). The analogous PMS high-altitude coronal structures would be associated with large-scale magnetic fields which are stronger on PMS stars that on the Sun, allowing the confinement of hotter plasma to higher altitudes.

Various theoretical models relevant to PMS stars support the observational evidence of enormous coronal magnetic structures to explain PMS X-ray super-flares.
\begin{itemize}

\item[--] \citet{Vidotto2009} develop a three-dimensional MHD simulation of a diskfree  young star that includes centrifugal, magnetic, gravitational, and thermal forces.  They predict the formation of large ($>5-10$~R$_{\star}$), stable, elongated magnetic streamers upon interaction of the stellar wind and magnetosphere assuming high ratios between the magnetic and thermal energy densities.  

\item [--] \citet{Aarnio2012} describe a model involving gravity, buoyancy, magnetic pressure gradient, magnetic tension, and stellar winds that show  hot post-flare coronal loops can acquire a mechanical equilibrium even at heights beyond the corotation radius. 

\item[--] \citet{Cohen2017} perform  magnetohydrodynamic calculations for a rapidly rotating, fully convective M star. Their simulations use a rotating spherical shell as a surrogate for a convection zone, generating both complex small-scale and dipole large-scale magnetic fields, combined with the Alfv\'{e}n wave turbulence to mimic coronal heating. The simulations exhibit large areas at high-latitude with a several kilo-Gauss surface magnetic fields that sustain strong high-altitude dipolar magnetic loops with confined hot plasma. 
\end{itemize}

These theoretical expectations support not only the estimations of large X-ray coronal loops on diskless COUP stars \citep{Getman08b} but interacting radio flaring streamers of sizes $>15$~$R_{\star}$ observed in the young, diskless binary system ($1.5$~M$_{\odot}$ and $1.3$~M$_{\odot}$) V773 Tau A \citep{Massi2008}. 

However, we recall our discussion in \S\ref{sec:flare_model} that the behavior of a flaring single loop on the solar surface can be mimicked by a flaring arcade of loops providing the triggering is nearly simultaneous. This possibility is particularly suggested by the unusually high loop thickness ($\beta$) values emerging from most of the PMS super-flare model fits (\S\ref{sec:thick_mass_dependence}). 

Finally, the underlying mechanism of the magnetic dynamo in PMS stars is thought to be fundamentally different from the solar dynamo.  PMS stars on the Hayashi track are fully convective and, even when slowed by star-disk coupling, rapidly rotating.  It is widely believed that such stars generate magnetic fields through distributed turbulent $\alpha^2$-type magnetic dynamos rather than tachoclinal $\alpha \Omega$-type dynamos  \citep{Durney1993, Yadav2015, Cohen2017, Warnecke2020}. Some calculations suggest that surface magnetic fields may be concentrated in high-latitude areas, but Zeeman effect observations of individual PMS stars indicate that complex multi-polar surface magnetic field configurations are typical \citep{Gregory2010}.

\subsection{On Star-Disk Magnetic Field Structures} \label{sec:stardisk_vs_starstar}

Since the first powerful flares on Class I protostars were detected in X-rays \citep{Koyama1996, Grosso1997}, theorists have calculated models of flares from magnetic field loops with one footprint in the stellar surface and the other footprint in the protoplanetary disk. The model has strong astrophysical motivation.  It has been convincingly argued that star-disk magnetic coupling slow the protostars' rapid rotation and regulate stellar angular momentum through the Class I and II accretion phases \citep{Bouvier1997}.  This torque is most readily explained by stellar magnetic field lines threading the disk and, since the disk is differentially rotating, shearing and reconnection of the star-disk field lines is a natural consequence. Models of such flares were first presented by \citet{Hayashi1996}; recent sophisticated time-dependent 3-dimensional MHD calculations of star-disk magnetic flaring have been calculated by \citet{Colombo2019} and \citet{Takasao2019}.  

However, our observational data do not provide any evidence for this process.  In Paper I, we show that disk-bearing PMS stars (including some protostars) and diskless PMS stars have indistinguishable X-ray luminosity and total energy distribution functions.  In the present paper, we show that flare properties $-$ such as the model-independent flare duration, peak emission measure, peak plasma temperature, and energy $-$ as well as the loop geometries inferred from a solar-like single loop model, are indistinguishable between disk-bearing and diskless stars (\S\ref{sec:disk_dependence}).  We further show here that, in most respects, PMS super-flares have properties that extend those of solar flares and older stellar flares where no disk is present (\S \ref{sec:solarlink},\ref{sec:solar_applicability}).

Our work supports studies by other groups. For instance, \citet{Aarnio2010} analyzed the infrared spectral energy distributions of the 32 COUP flare hosts from \citet{Favata2005} showing that many long X-ray loops are associated with diskless stars. \citet{LopezSantiago2010} derive an extremely long flaring loop for the $8$~Myr old, diskless star TWA 11B. The main difference between diskless and disk-bearing PMS stars is a suppression of `characteristic' X-ray emission in accreting stars \citep{Preibisch05} that can be explained e.g., by soft X-ray absorption in accretion columns \citep{Gregory2007}. 

We see three ways to reconcile the need for star-disk magnetic coupling with the observational evidence.  First, the properties of flaring with star-disk footprints and star-star footprints are nearly identical, despite the different geometries and causes of reconnection. Second, the star-disk reconnection events are frequent and weak, so their X-ray emission buried in the continuous `characteristic' emission of PMS stars.  Third,  the star-disk flares occur much more rarely than star-star flares and have not yet been detected. 

We suggest the second explanation is plausible: when differential rotation between the star and disk stretches and breaks a magnetic field line, the resulting field lines could be open rather than closed. Energetic particles generated in the reconnection event would not efficiently evaporate gas from the disk or stellar surface. Any plasma generated would be unconfined, not attaining the high densities needed for substantial X-ray emission.


\subsection{Dependence on Stellar Mass} \label{sec:disucssion_mass_dependence}

Section \S\ref{sec:coup2} presents positive correlations between stellar mass and the super-flare peak X-ray power and peak temperature.  Relationships between the characteristic X-ray luminosity are mass have been well-established in studies of the Orion Nebula Cluster and Taurus PMS stars \citep{Preibisch05,Telleschi07}. Here the X-ray luminosity is either averaged across long X-ray observations or obvious large X-ray flares are excluded. This characteristic X-ray emission is likely the product of numerous superposed weaker flares \citep{Wolk05}. The extension of these relationships, from weaker flares to super-flares points towards common physics among a wide range of PMS flaring behavior. 

 We note that since stellar mass, radius, surface area and volume are mutually correlated along the Hayashi tracks in Hertzsprung-Russell diagram, it is difficult to distinguish the underlying physical cause of this relationship.  For example, it is plausible that the most relevant property is surface area that scales with the magnetic flux and X-ray luminosity in the Sun and stars (Figure~\ref{fig:pevtsov_plot}, \S\ref{sec:solarlink}). 

However, a relationship between the loop thickness parameter $\beta$ and stellar mass is newly reported here  \S\ref{sec:thick_mass_dependence}.  We can tentatively exclude some reasons for this effect.  First, \citet{Sokal2020} do not find a relationship between surface magnetic field and stellar mass among PMS stars.  Second, Figure~\ref{fig:bright_superflares_mass_strata} shows no increase in loop height with stellar mass.  Third,  Table~\ref{tab:55_brightest} indicates wide ranges of plasma densities ($n=10^{10}-10^{12.5}$~cm$^{-3}$) in the super-flaring loops with no differences for higher- and low-mass stars.

The remaining cause of increasing $\beta$ in loops with geometries associated with higher volume, either due to thicker single loops or to loop arcades, or possibly inhomogeneities in density. We encourage MHD  calculations of dynamos in fully convective stars to examine a range of stellar masses to see if the magnetic configuration at the stellar surface or in the loop changes with increasing stellar mass.

\section{Conclusions} \label{sec:conclusions}

In the present study, we examine a subset of 55 MYStIX/SFiNCs  super-flares captured in $>1000$-photon {\it Chandra} observations, sufficiently  bright  to  analyze  the temporal history of the flares in detail. Temporal profiles of various flare properties $-$ such as emission measure, plasma temperature, and cooling time-scales $-$ are produced and show the similar multi-phase heating and cooling behaviors seen in single loop solar flares.  We are thereby motivated, like many previous studies, to model the flares using the solar loop model of \citet{Reale1997} and estimate flare geometries (\S \ref{sec:methods2}).

These MYStIX/SFiNCs flares, combined with the previously analysed COUP flares \citep{Favata2005,Getman08a}, form the largest sample of most energetic PMS X-ray flares ever modelled. We find that the variety of their lightcurve morphology is not reflected in their peak X-ray luminosity or plasma temperature (\S \ref{sec:variety}). But clear patterns are seen between flare properties and stellar mass. Super-flare processes in more massive PMS stars are often governed by conduction cooling (\S \ref{sec:cooling_vs_plasma_temperature}), occur in thicker coronal flaring structures, and are likely associated with larger surface star spots (\S  \ref{sec:thick_mass_dependence}, \ref{sec:disucssion_mass_dependence}). 

The properties of PMS super-flares are independent from the presence or absence of protoplanetary disks, supporting the solar-type model of PMS flaring magnetic loops with both footpoints anchored in the stellar surface (\S \ref{sec:disk_dependence}, \ref{sec:stardisk_vs_starstar}). 

The joint MYStIX/SFiNCs-COUP super-flare sample shows that the peak flare emission measure, total flare energy, and flare peak plasma temperature scale with stellar mass (\S \ref{sec:coup2}). This is an extension of similar relationships seen for the produce of numerous superposed weaker flares pointing towards common physics among a wide range of PMS flaring behavior (\S \ref{sec:disucssion_mass_dependence}). 

PMS  super-flares  with  their long coronal loops show some common properties to rare long-duration eruptive solar flares that are accompanied by large-scale coronal mass ejections (\S \ref{sec:older_stars2}, \ref{sec:solar_applicability}).  In the Sun, they carry far more energy in particles and magnetic fields into the environment than is radiated by the flare itself.  We encourage observational efforts to detect the elusive signatures of coronal mass ejections in flaring PMS stars.  

Finally, the slope of a long-standing relationship between the X-ray luminosity and magnetic flux of various solar-stellar magnetic elements appears steeper in PMS super-flares than for solar events (\S \ref{sec:solarlink}). This requires surface field strengths where the flare footprints are lodged to be considerably stronger, perhaps $B > 20$~kG, than seen on the Sun.   

\acknowledgments
We are grateful to the referee for providing useful suggestions that stimulated fresh ideas and improved the paper. This project is supported by the {\it Chandra} archive grant AR9-20002X and the {\it Chandra} ACIS Team contract SV474018 (G. Garmire \& L. Townsley, Principal Investigators), issued by the {\it Chandra} X-ray Center, which is operated by the Smithsonian Astrophysical Observatory for and on behalf of NASA under contract NAS8-03060. The {\it Chandra} Guaranteed Time Observations (GTO) data used here were selected by the ACIS Instrument Principal Investigator, Gordon P. Garmire, of the Huntingdon Institute for X-ray Astronomy, LLC, which is under contract to the Smithsonian Astrophysical Observatory; contract SV2-82024. This research has made use of NASA's Astrophysics Data System Bibliographic Services. 

\vspace{5mm}
\facilities{CXO}

\software{ACIS Extract \citep{Broos10}, 
        R \citep{RCoreTeam20}}

\clearpage
\appendix

\section{X-ray Photon Arrival Diagrams} \label{sec:photon_arrival}
\begin{figure*}[ht]
\epsscale{1.2}
\plotone{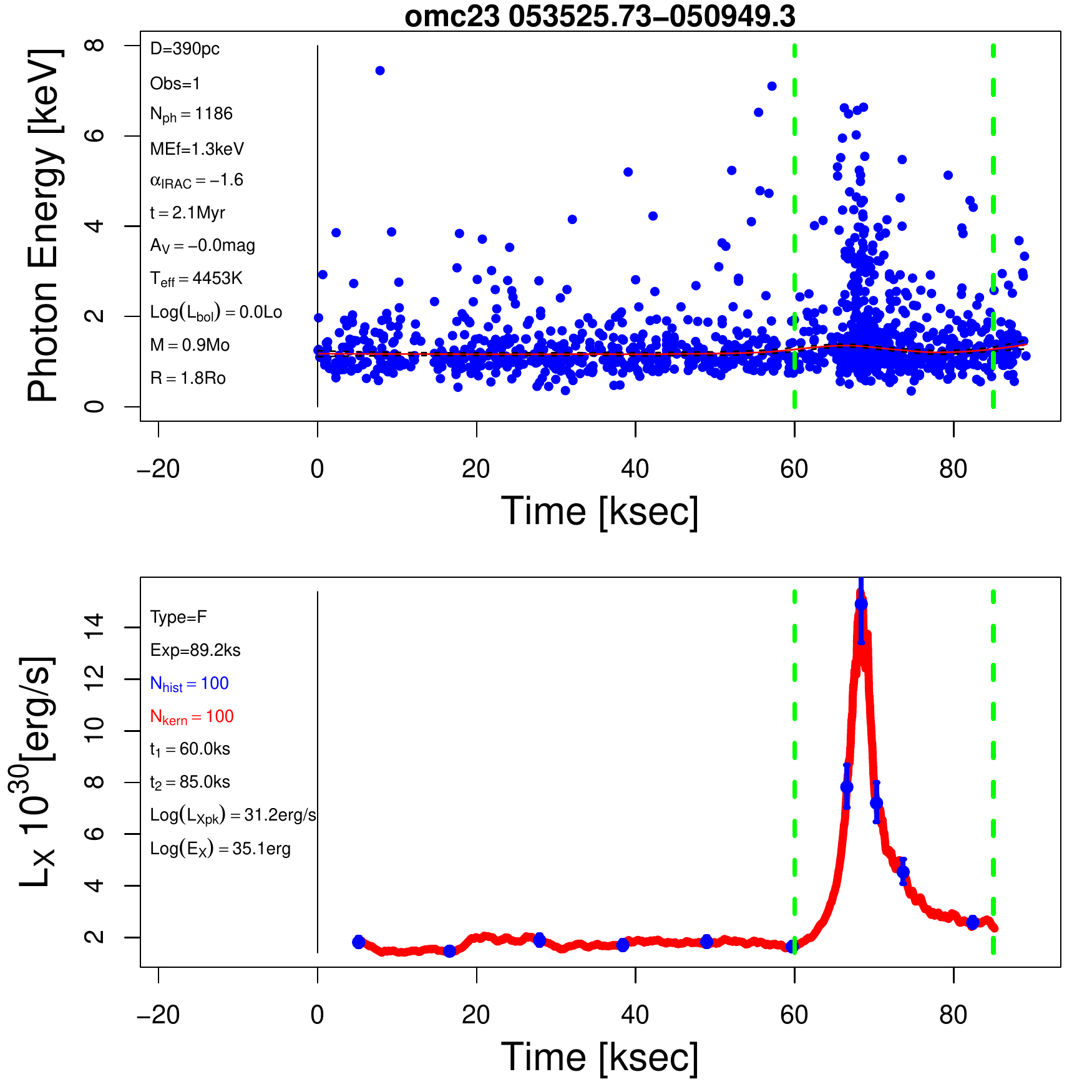}
\caption{Sample  page  from  the photon arrival atlas for our modelled flare \#1 from Figure \ref{fig:atlas_example1}. See text for details.   The complete figure set (55 images) is available in the electronic version of the paper. \label{fig:photon_arrival}}
\end{figure*}

Figure \ref{fig:photon_arrival} presents an example of the photon arrival diagram and X-ray lightcurve for our super-flare \#1, spanning the entire duration of the flare-host {\it Chandra} observation. The electronic atlas associated with Figure~\ref{fig:photon_arrival} provides such information for all 55 bright super-flares examined in this study. A similar in format flare atlas for all 1086 MYStIX/SFiNCs super-flares is available as an electronic figure set in Paper I.

This figure features two plots with the host star and flare properties extracted from Tables given in Paper I.  The first plot gives X-ray photon arrival times and energies with individual photons marked as blue points.  The red curve shows a likelihood-based local quadratic regression fit with 84\% CIs (black dashed lines) generated using the {\it locfit.robust} function from CRAN package {\rm locfit} package \citep{Loader20}. This procedure and its mathematical foundations are described by \citet{Loader99}.  The flare start/stop times indicated by the green dashed lines. 

The plot title gives the star identifier and its star formation region. The annotation gives 11 scalar quantities: distance to the star formation region; relative number of the current {\it Chandra} ObsID for this X-ray source; number of total X-ray photons in the ObsID with the flare ($N_{phot}$); median energy of the ObsID; infrared slope from which disk presence is inferred; estimated age, visual absorption, stellar effective temperature, bolometric luminosity, mass and radius derived using X-ray-IR photometry data as described in Paper I. 

The $N_{phot} > 1000$ photons criterion used in the current paper to select a sample of 55 bright MYStIX/SFiNCs super-flares is based on the assumption that most of the super-flares dominate the total counts of their host {\it Chandra} observations. This assumption does not apply to all 1086 MYStIX/SFiNCs super-flares from Paper I; and hence, some of bright super-flares within {\it Chandra} exposures with total counts $N_{phot} < 1000$ are excluded from the modeling analysis of the current paper. Specifically, among the 55 bright super-flares modelled here, flare \#30 (m17 182028.36-161030.6\_2) has the lowest number of total X-ray counts within the start/stop flare times (green dashed lines in Figure~\ref{fig:photon_arrival}), $N_{t1,t2}$ = 454 counts\footnote{Notice that the $N_{t1,t2}$ quantity is often higher than the $N_{b-g}$ quantity given in Figure \ref{fig:atlas_example1}. The latter indicates the number of X-ray counts in the modelled portion of a flare only.}. Among the 1031 MYStIX/SFiNCs super-flares not included in the current flare modeling analysis, 47 flares have $N_{t1,t2} > 454$ counts.       

The second plot gives an adaptively smoothed absorption-corrected X-ray luminosity lightcurve (red) with 1-$\sigma$ confidence intervals (CIs) for a binned histogram (blue) using analytical approximations for a Poisson distribution \citep{Gehrels86}. The binned histogram is composed of independent count bins, each accumulating similar numbers of X-ray counts ($N_{hist}$) and centered at the mean arrival time between the first and the last counts in the bin. The plot legends include: flare type;  ObsID exposure time; number of counts per adaptive kernel (red curve); number of counts in a histogram bin (blue); flare start time; flare stop time; $\log L_{X,pk}$; and $\log E_X$. Notice that the peak X-ray luminosity value here can be slightly different from that given in Figure \ref{fig:atlas_example1} due to different choices of an adaptive kernel. On the X-ray luminosity time series plots, not all jiggles in the red curve are statistically significant, and blue circles-with-errors are not carefully placed with respect to possible interesting structures such as flare peaks.

\section{Statistical Issues in Estimating Flare Parameters} \label{sec:mle}

\subsection{Rise and Decay Timescales}

It is well-known that the lightcurves of solar and stellar flares can be modelled using the exponential function. For instance, for powerful COUP flares \citet{Favata2005, Getman08a} estimate flare e-folding rise ($\tau_{rise}$) and decay ($\tau_{decay}$) timescales by fitting the observed binned (with independent bins) count rate ($CR$) time series with exponential rise and decay models (using least squares on $ln(CR)$):
\begin{equation} \label{eqn:cr_vs_t}
CR=CR_{flare,pk} \times \exp(\frac{t - t_2}{\tau})
\end{equation} 
where $t_2 = t_{CR,pk}$ is the time of the peak count rate, and $\tau$ is $\tau_{rise}$ or $-\tau_{decay}$.  

For the decay phase of a flare, by integrating over a time range and normalizing to the total X-ray counts, this function can be transformed to the cumulative distribution function (CDF) of unbinned (raw) X-ray count arrival times
\begin{equation} \label{eq:CDF_exponetial}
CDF~  =  \frac{1-exp(-\lambda t)}{C_{\lambda}, } \nonumber
\end{equation}
\begin{equation} \label{eq:coeff_exponetial}
{\rm where~~} C_{\lambda} = 1 - exp(-\lambda t_{m})
\end{equation}
\begin{equation} \label{eq:range_exponetial}
t~ \epsilon~ [0,t_{m}]  \nonumber
\end{equation}

\begin{figure*}[ht]
\epsscale{1.2}
\plotone{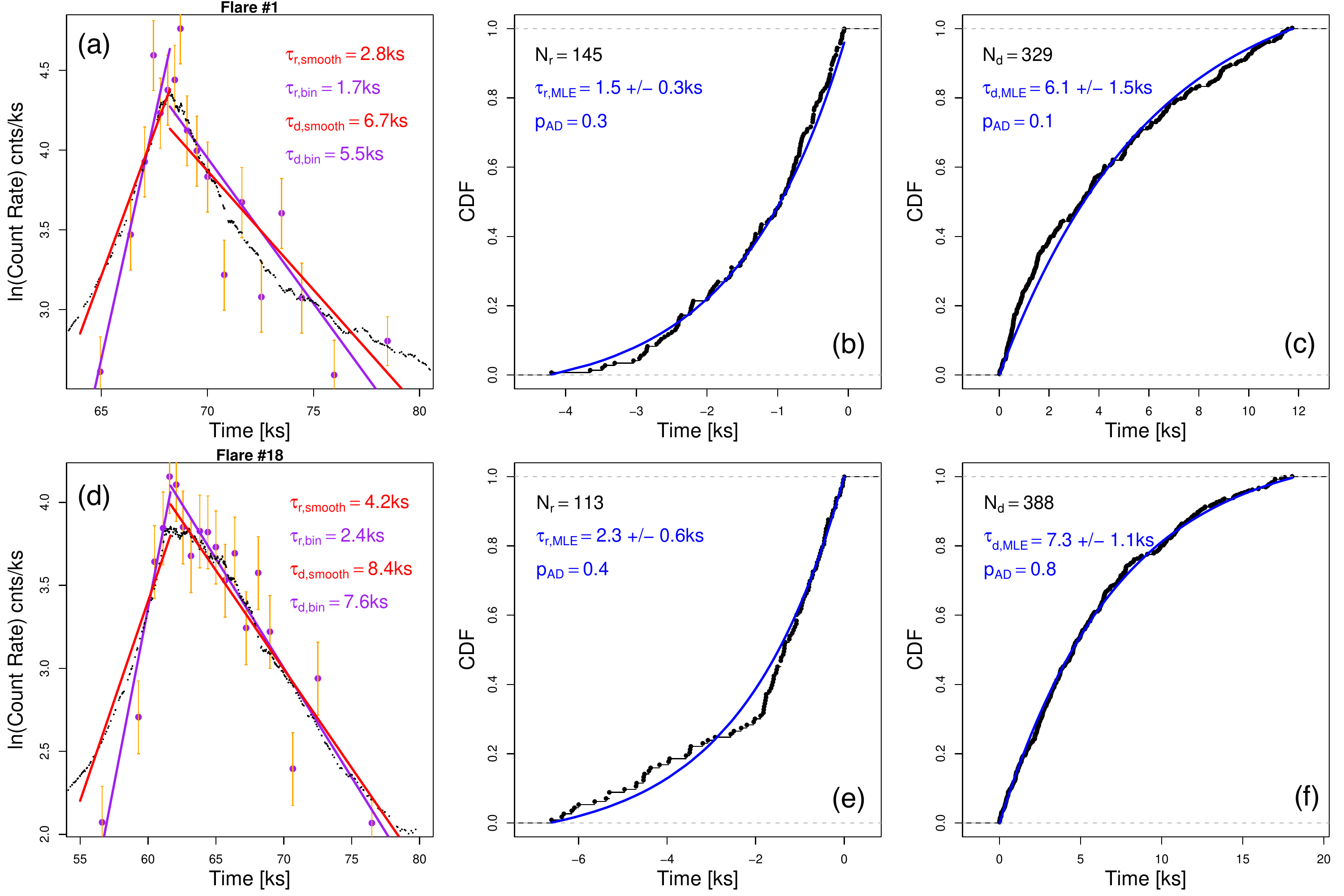}
\caption{Rise and decay timescales for the flares from Figures \ref{fig:atlas_example1} and \ref{fig:atlas_example2}. (a,d) X-ray lightcurves. Adaptively smoothed versions ($N_{kern} = [130-150]$ counts) are indicated by black points. Independent histogram bins ($N_{bin} = 25$ counts) are shown as purple points with error bars (orange) as approximations to 1-$\sigma$ CIs of a Poissonian distribution \citep{Gehrels86}. Linear rise/decay fits for arrival times, using unweighted and weighted least squares for smoothed and binned data, are shown as red and purple lines, respectively. The inferred time-scales are labelled. (b,c,e,f) Cumulative distribution functions of the arrival times of raw X-ray counts (black points) with the best fits using the maximum likelihood estimator (blue). (b,e) and (c,f) panels present the flare rise and decay phases, respectively, with the time point at $0$~ks referencing the instant of maximum smoothed count rate. The legends list the numbers of X-ray photons, inferred timescales using MLE, their 95\% CIs from 1000 bootstrap re-samples, and $p$-values from Anderson-Darling goodness-of-fit tests. \label{fig:mle_example}}
\end{figure*}

\noindent Here, time is referenced to the observed moment with maximum count rate ($t_{maxCR}$), so that the minimum and maximum time points in the aforementioned equations are $t_{maxCR}-t_{maxCR} = 0$ and $t_{m} = t_{maxDecay} - t_{maxCR}$, respectively, where $t_{maxDecay}$ is the observed stop time of the decay phase. The parameter $\lambda$ is related to the e-folding decay timescale of the flare lightcurve as $\tau_{decay} = 1/\lambda$. As an example, for the decay phases of the flares from Figures \ref{fig:atlas_example1} and \ref{fig:atlas_example2}, the fits of such a CDF model (blue) to the raw data (black) are shown in Figures \ref{fig:mle_example}(c,f). The related probability density function (PDF) and likelihood function ($L$) can be derived
\begin{equation} \label{eq:PDF_exponetial}
PDF =  \frac{dCDF}{dt} = \frac{\lambda exp(-\lambda t)}{C_{\lambda}} \nonumber 
\end{equation}
\begin{equation} \label{eq:L_exponetial}
{\rm and~~}ln(L) = ln(\prod_{1}^{N} PDF(t_{i})) = N ln(\lambda/C_{\lambda}) - \lambda N \hat{t},
\end{equation}
\begin{equation} \label{eq:meant_exponential}
{\rm where~~}\hat{t} = \frac{1}{N} \sum_{1}^{N} t_{i}
\end{equation}

\noindent and the maximum likelihood estimator $\hat{\lambda}$ can be found by numerically solving

\begin{equation} \label{eq:mle_exponetial}
\frac{dln(L)}{d\lambda} =  \frac{N}{\hat{\lambda}} - \frac{N t_{m} exp(-\hat{\lambda} t_{m})}{C_{\hat{\lambda}}} - N \hat{t} = 0
\end{equation}

For the rise phase of a flare, the above equations remain the same except the time range should be substituted with $[t_{minRise} - t_{maxCR}, 0]$, where $t_{minRise}$ is the observed start of the rise phase. For such rise phase range, the cumulative distribution of raw data is fit by the $1 - CDF$ function from equation \ref{eq:CDF_exponetial}, and the e-folding rise timescale of the flare lightcurve is $\tau_{rise} = - 1/\lambda$. For the flares from Figures \ref{fig:atlas_example1} and \ref{fig:atlas_example2}, the maximum likelihood estimator (MLE) of their rise phases are shown in Figures \ref{fig:mle_example}(b,e).

In the panel (c) of the flare atlas (Figure Set \ref{fig:atlas_example1}; \S \ref{sec:masme_method}), the MLE timescale estimates on the raw data are listed in the panel legends, and the corresponding model fits are shown as the blue and green curves for the rise and decay flare phases, respectively. The loop geometries ($L_{decay}, \beta$), ($L_{rise}$), and cooling timescales described in \S \ref{sec:flare_model} and \ref{sec:cooling} are calculated using these timescales. These MLE timescales derived from unbinned data along with their 95\% CIs are further reported in Table~\ref{tab:55_brightest}.

But two other statistical approaches can be considered. One is to fit the exponential function to the adaptively smoothed X-ray time series.  There is a delicate balance between under-smoothing that would insufficiently suppress noise, and over-smoothing that could miss important global features. We have investigated a smoothing kernel starting with a large kernel size of $N_{kern} = 300$ X-ray counts and iteratively adjusting the kernel size until the flare e-folding decay timescale calculated on smoothed data, using equation \ref{eqn:cr_vs_t}, falls within the 95\% CI of the decay timescale for unbinned data. While smoothing has the advantage of reducing noise, it has the disadvantage that adjacent points are no longer independent.  This failure of the independence assumption of a least squares regression may lead to bias in the timescale estimates, and their uncertainties.  

A binned approach employed by \citet{Favata2005} and \citet{Getman08a}  has other difficulties.  The arbitrary choice of bin width and origin violates the theorem by \citet{Chernoff54} who show that the degrees of freedom are ill-defined in such situation. Again the resulting timescale uncertainties may not be reliable. 

For these reasons, in Table~\ref{tab:55_brightest} we report only flare timescales derived using maximum likelihood estimation from the unbinned X-ray arrival times. However, we have examined both the binned and smoothed approaches and find little difference in the estimation of decay timescales, with small biases and spreads of $<20$\% comparable to the statistical errors on the MLE timescales obtained from raw data. The smoothed rise timescales remain systematically higher (by about $40$\%) than the binned or raw timescales indicating that the rise phases for our modelled flares could be oversmoothed.

\subsection{Temperature-Density Slope}

Estimation of the slope $\zeta$ in the $T - \sqrt{EM}$ diagram is more uncertain.  First, as these quantities can only be estimated from ensembles of photons, smoothing or binning is necessary. In the flare atlas (Figure Set~\ref{fig:atlas_example1}), we report slopes from adaptively smoothed lightcurves.  Second, the observed pattern of points is sometimes not linear; in statistical parlance, the linear model is misspecified. For example, the green points in the lower-left panels of flares in Figures~\ref{fig:atlas_example1} and \ref{fig:atlas_example2} both show a convex curvature with respect to the green lines.  This indicates the flare plasma is cooling faster, and other times slower, during portions of the decay compared to the power law behavior expected from the \citet{Reale1997} model.  

\begin{figure*}[ht]
\epsscale{1.0}
\plotone{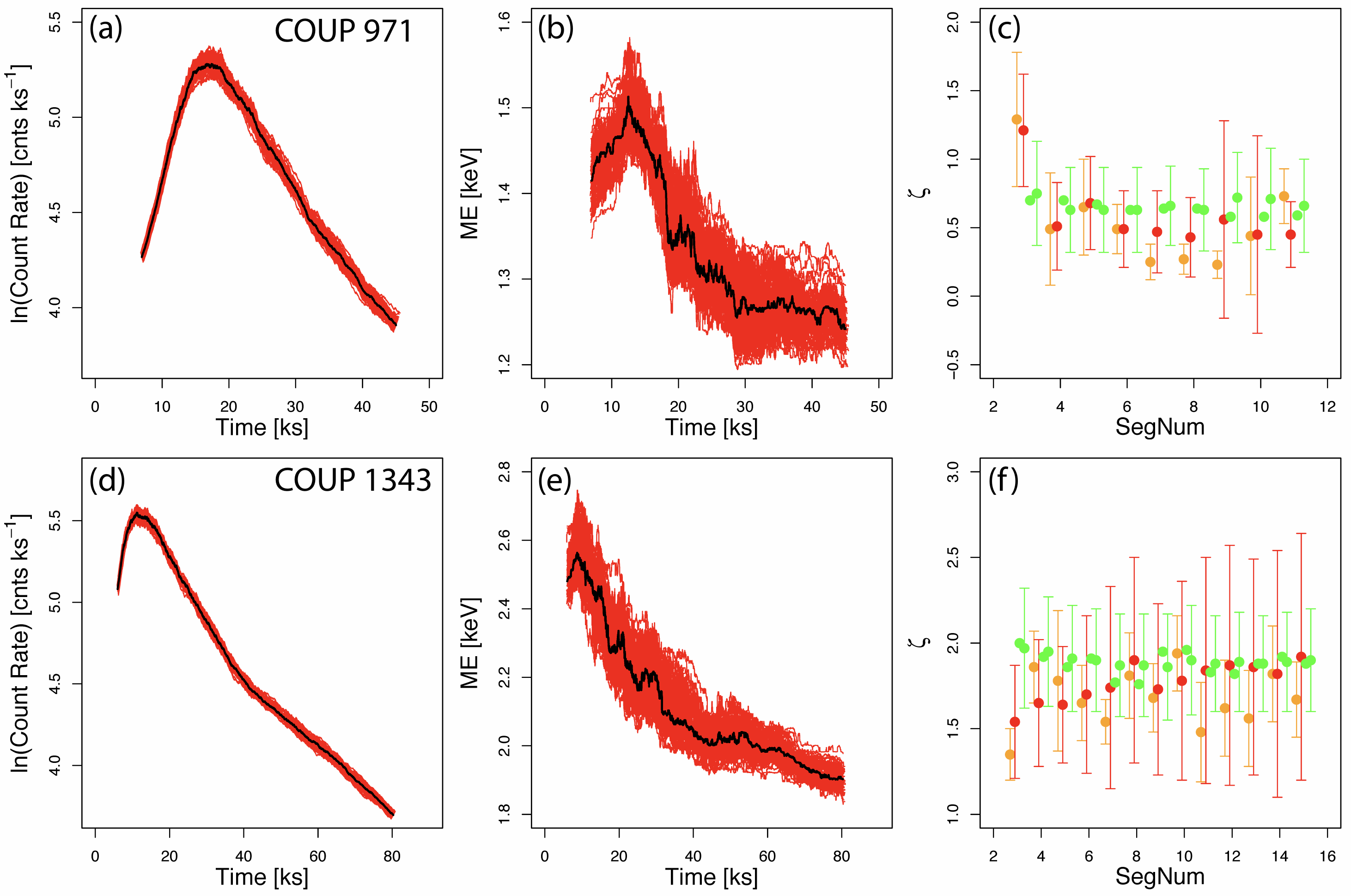}
\caption{Results from the tests using re-sampling of the original {\it Chandra} X-ray photons for the COUP 971 (a,b,c) and COUP 1343 (d,e,f) flares. (a,b,d,e) Temporal evolution of adaptively smoothed count rate and median energy for the original 4957-count and 9602-count data of the COUP 971 and 1343 flares, respectively (black curves). One hundred re-samples for each of the flares from the original X-ray photons (red loci). (c,f) Inferred slope $\zeta$ for the temporal flare evolution on the temperature-density diagram as a function of the number of independent data segments ($SegNum$) within the decay phases of the flares. Results from the linear regression fits of the smoothed and binned data for the original (not re-sampled) flares are shown as green points without error bars and orange points with 1-$\sigma$ error bars, respectively. The mean and standard deviation for the $\zeta$ distributions derived from 100 flare re-samples are shown for the smoothed and binned data versions as green points with error bars and red points with error bars, respectively. \label{fig:slope_stability}}
\end{figure*}

To help evaluate the reliability of the linear fit estimates of $\zeta$, we have conducted tests involving bootstrap re-sampling  of the original X-ray photons during the decay phases of two rich COUP flares, COUP 971 and COUP 1343, which exhibit different cooling behaviours with $\zeta$ values around 0.7 and 1.8, respectively \citep{Getman08a}. Both flares are subject to minor reheating processes during flare decay stages.  The effects of different choices of smoothing and binning are taken into consideration\footnote{The data are adaptevely smoothed using kernel size $N_{kern}$ and are binned providing independent bins with size $N_{hist}$. For COUP 971, the analyses are performed for different sets of kernel sizes ($N_{kern} = N_{hist} = [$ 300, 310, 320, 380, 400, 500, 600, 700, 1000 $]$ counts) corresponding to the following numbers of independent bins (in case of binned data) along the entire decay phase of the flare ($SegNum = [11, 10, 9, 8, 7, 6, 5, 4, 3]$). For COUP 1343, the sizes are $N_{kern} = N_{hist} = [$ 500, 600, 640, 700, 715, 800, 830, 1000, 1100, 1200, 1500, 1700, 2000 $]$ counts, corresponding to the numbers of independent bins $SegNum = [15, 14, 13, 12, 11, 10, 9, 8, 7, 6, 5, 4, 3]$).}. Figure \ref{fig:slope_stability} shows that $\zeta$ based on smoothed data of the original (green points without error bars) and re-sampled (green points with error bars) flares are much more statistically stable than $\zeta$ derived from binned data of the original (orange points with error bars) and re-sampled (red points with error bars) flares. These tests suggest that use of smoothed data is preferred over that of binned data. 

Altogether, we believe that the reported $\zeta$ values based on smoothed data are reasonably reliable within the assumption that the cooling behavior follows the \citet{Reale1997} model.  But when the flare cooling does not trace a straight line in the $T - \sqrt{EM}$ diagram, the meaning of $\zeta$ is less clear.  Similar nonlinear cooling behaviors are seen in ONC super-flares \citep{Getman08a}.  More sophisticated astrophysical cooling models are needed for these super-flares.

\bibliography{my_bibliography}{}

\begin{thebibliography}{}
\expandafter\ifx\csname natexlab\endcsname\relax\def\natexlab#1{#1}\fi
\providecommand{\url}[1]{\href{#1}{#1}}
\providecommand{\dodoi}[1]{doi:~\href{http://doi.org/#1}{\nolinkurl{#1}}}
\providecommand{\doeprint}[1]{\href{http://ascl.net/#1}{\nolinkurl{http://ascl.net/#1}}}
\providecommand{\doarXiv}[1]{\href{https://arxiv.org/abs/#1}{\nolinkurl{https://arxiv.org/abs/#1}}}

\bibitem[{{Aarnio} {et~al.}(2012){Aarnio}, {Llama}, {Jardine}, \&
  {Gregory}}]{Aarnio2012}
{Aarnio}, A., {Llama}, J., {Jardine}, M., \& {Gregory}, S.~G. 2012, \mnras,
  421, 1797, \dodoi{10.1111/j.1365-2966.2012.20434.x}

\bibitem[{{Aarnio} {et~al.}(2010){Aarnio}, {Stassun}, \& {Matt}}]{Aarnio2010}
{Aarnio}, A.~N., {Stassun}, K.~G., \& {Matt}, S.~P. 2010, \apj, 717, 93,
  \dodoi{10.1088/0004-637X/717/1/93}

\bibitem[{{Albacete Colombo} {et~al.}(2007){Albacete Colombo}, {Caramazza},
  {Flaccomio}, {Micela}, \& {Sciortino}}]{Colombo07}
{Albacete Colombo}, J.~F., {Caramazza}, M., {Flaccomio}, E., {Micela}, G., \&
  {Sciortino}, S. 2007, \aap, 474, 495, \dodoi{10.1051/0004-6361:20078064}

\bibitem[{{Argiroffi} {et~al.}(2016){Argiroffi}, {Caramazza}, {Micela},
  {Sciortino}, {Moraux}, {Bouvier}, \& {Flaccomio}}]{Argiroffi2016}
{Argiroffi}, C., {Caramazza}, M., {Micela}, G., {et~al.} 2016, \aap, 589, A113,
  \dodoi{10.1051/0004-6361/201526539}

\bibitem[{{Argiroffi} {et~al.}(2006){Argiroffi}, {Favata}, {Flaccomio},
  {Maggio}, {Micela}, {Peres}, \& {Sciortino}}]{Argiroffi2006}
{Argiroffi}, C., {Favata}, F., {Flaccomio}, E., {et~al.} 2006, \aap, 459, 199,
  \dodoi{10.1051/0004-6361:20065674}

\bibitem[{{Aschwanden} \& {Alexander}(2001)}]{Aschwanden2001}
{Aschwanden}, M.~J., \& {Alexander}, D. 2001, \solphys, 204, 91,
  \dodoi{10.1023/A:1014257826116}

\bibitem[{{Aschwanden} {et~al.}(2000){Aschwanden}, {Nightingale}, \& {Alexand
  er}}]{Aschwanden2000}
{Aschwanden}, M.~J., {Nightingale}, R.~W., \& {Alexand er}, D. 2000, \apj, 541,
  1059, \dodoi{10.1086/309486}

\bibitem[{{Aschwanden} {et~al.}(2008){Aschwanden}, {Stern}, \&
  {G{\"u}del}}]{Aschwanden2008}
{Aschwanden}, M.~J., {Stern}, R.~A., \& {G{\"u}del}, M. 2008, \apj, 672, 659,
  \dodoi{10.1086/523926}

\bibitem[{{Benz}(2017)}]{Benz2017}
{Benz}, A.~O. 2017, Living Reviews in Solar Physics, 14, 2,
  \dodoi{10.1007/s41116-016-0004-3}

\bibitem[{{Bhatt} {et~al.}(2014){Bhatt}, {Pandey}, {Singh}, {Sagar}, \&
  {Kumar}}]{Bhatt2014}
{Bhatt}, H., {Pandey}, J.~C., {Singh}, K.~P., {Sagar}, R., \& {Kumar}, B. 2014,
  Journal of Astrophysics and Astronomy, 35, 39,
  \dodoi{10.1007/s12036-014-9297-6}

\bibitem[{{Bouvier} {et~al.}(1997){Bouvier}, {Forestini}, \&
  {Allain}}]{Bouvier1997}
{Bouvier}, J., {Forestini}, M., \& {Allain}, S. 1997, \aap, 326, 1023

\bibitem[{{Bressan} {et~al.}(2012){Bressan}, {Marigo}, {Girardi}, {Salasnich},
  {Dal Cero}, {Rubele}, \& {Nanni}}]{Bressan12}
{Bressan}, A., {Marigo}, P., {Girardi}, L., {et~al.} 2012, \mnras, 427, 127,
  \dodoi{10.1111/j.1365-2966.2012.21948.x}

\bibitem[{{Broos} {et~al.}(2010){Broos}, {Townsley}, {Feigelson}, {Getman},
  {Bauer}, \& {Garmire}}]{Broos10}
{Broos}, P.~S., {Townsley}, L.~K., {Feigelson}, E.~D., {et~al.} 2010, \apj,
  714, 1582, \dodoi{10.1088/0004-637X/714/2/1582}

\bibitem[{{Caramazza} {et~al.}(2007){Caramazza}, {Flaccomio}, {Micela},
  {Reale}, {Wolk}, \& {Feigelson}}]{Caramazza07}
{Caramazza}, M., {Flaccomio}, E., {Micela}, G., {et~al.} 2007, \aap, 471, 645,
  \dodoi{10.1051/0004-6361:20077195}

\bibitem[{{Cargill} {et~al.}(1995){Cargill}, {Mariska}, \&
  {Antiochos}}]{Cargill1995}
{Cargill}, P.~J., {Mariska}, J.~T., \& {Antiochos}, S.~K. 1995, \apj, 439,
  1034, \dodoi{10.1086/175240}

\bibitem[{{Chen} {et~al.}(2014){Chen}, {Girardi}, {Bressan}, {Marigo},
  {Barbieri}, \& {Kong}}]{Chen14}
{Chen}, Y., {Girardi}, L., {Bressan}, A., {et~al.} 2014, \mnras, 444, 2525,
  \dodoi{10.1093/mnras/stu1605}

\bibitem[{Chernoff \& Lehmann(1954)}]{Chernoff54}
Chernoff, H., \& Lehmann, E.~L. 1954, The Annals of Mathematical Statistics,
  25, 579 , \dodoi{10.1214/aoms/1177728726}

\bibitem[{{Cohen} {et~al.}(2017){Cohen}, {Yadav}, {Garraffo}, {Saar}, {Wolk},
  {Kashyap}, {Drake}, \& {Pillitteri}}]{Cohen2017}
{Cohen}, O., {Yadav}, R., {Garraffo}, C., {et~al.} 2017, \apj, 834, 14,
  \dodoi{10.3847/1538-4357/834/1/14}

\bibitem[{{Colombo} {et~al.}(2019){Colombo}, {Orlando}, {Peres}, {Reale},
  {Argiroffi}, {Bonito}, {Ibgui}, \& {Stehl{\'e}}}]{Colombo2019}
{Colombo}, S., {Orlando}, S., {Peres}, G., {et~al.} 2019, \aap, 624, A50,
  \dodoi{10.1051/0004-6361/201834342}

\bibitem[{{Durney} {et~al.}(1993){Durney}, {De Young}, \&
  {Roxburgh}}]{Durney1993}
{Durney}, B.~R., {De Young}, D.~S., \& {Roxburgh}, I.~W. 1993, \solphys, 145,
  207, \dodoi{10.1007/BF00690652}

\bibitem[{{Favata} {et~al.}(2005){Favata}, {Flaccomio}, {Reale}, {Micela},
  {Sciortino}, {Shang}, {Stassun}, \& {Feigelson}}]{Favata2005}
{Favata}, F., {Flaccomio}, E., {Reale}, F., {et~al.} 2005, \apjs, 160, 469,
  \dodoi{10.1086/432542}

\bibitem[{{Favata} \& {Schmitt}(1999)}]{Favata1999}
{Favata}, F., \& {Schmitt}, J.~H.~M.~M. 1999, \aap, 350, 900.
\newblock \doarXiv{astro-ph/9909041}

\bibitem[{{Feigelson} {et~al.}(2013){Feigelson}, {Townsley}, {Broos}, {Busk},
  {Getman}, {King}, {Kuhn}, {Naylor}, {Povich}, {Baddeley}, {Bate},
  {Indebetouw}, {Luhman}, {McCaughrean}, {Pittard}, {Pudritz}, {Sills}, {Song},
  \& {Wadsley}}]{Feigelson13}
{Feigelson}, E.~D., {Townsley}, L.~K., {Broos}, P.~S., {et~al.} 2013, \apjs,
  209, 26, \dodoi{10.1088/0067-0049/209/2/26}

\bibitem[{{Fisher} {et~al.}(1998){Fisher}, {Longcope}, {Metcalf}, \&
  {Pevtsov}}]{Fisher1998}
{Fisher}, G.~H., {Longcope}, D.~W., {Metcalf}, T.~R., \& {Pevtsov}, A.~A. 1998,
  \apj, 508, 885, \dodoi{10.1086/306435}

\bibitem[{{Flaccomio} {et~al.}(2018){Flaccomio}, {Micela}, {Sciortino}, {Cody},
  {Guarcello}, {Morales-Calder{\`o}n}, {Rebull}, \& {Stauffer}}]{Flaccomio2018}
{Flaccomio}, E., {Micela}, G., {Sciortino}, S., {et~al.} 2018, \aap, 620, A55,
  \dodoi{10.1051/0004-6361/201833308}

\bibitem[{{Franciosini} {et~al.}(2007){Franciosini}, {Pillitteri}, {Stelzer},
  {Micela}, {Briggs}, {Scelsi}, {Telleschi}, {Audard}, {Palla}, \&
  {G{\"u}del}}]{Franciosini2007}
{Franciosini}, E., {Pillitteri}, I., {Stelzer}, B., {et~al.} 2007, \aap, 468,
  485, \dodoi{10.1051/0004-6361:20066536}

\bibitem[{{Gehrels}(1986)}]{Gehrels86}
{Gehrels}, N. 1986, \apj, 303, 336, \dodoi{10.1086/164079}

\bibitem[{{Getman} {et~al.}(2011){Getman}, {Broos}, {Salter}, {Garmire}, \&
  {Hogerheijde}}]{Getman2011}
{Getman}, K.~V., {Broos}, P.~S., {Salter}, D.~M., {Garmire}, G.~P., \&
  {Hogerheijde}, M.~R. 2011, \apj, 730, 6, \dodoi{10.1088/0004-637X/730/1/6}

\bibitem[{{Getman} \& {Feigelson}(2021)}]{Getman2021}
{Getman}, K.~V., \& {Feigelson}, E.~D. 2021, ApJ in press, arXiv e-prints,
  arXiv:2105.04768.
\newblock \doarXiv{2105.04768}

\bibitem[{{Getman} {et~al.}(2008{\natexlab{a}}){Getman}, {Feigelson}, {Broos},
  {Micela}, \& {Garmire}}]{Getman08a}
{Getman}, K.~V., {Feigelson}, E.~D., {Broos}, P.~S., {Micela}, G., \&
  {Garmire}, G.~P. 2008{\natexlab{a}}, \apj, 688, 418, \dodoi{10.1086/592033}

\bibitem[{{Getman} {et~al.}(2010){Getman}, {Feigelson}, {Broos}, {Townsley}, \&
  {Garmire}}]{Getman10}
{Getman}, K.~V., {Feigelson}, E.~D., {Broos}, P.~S., {Townsley}, L.~K., \&
  {Garmire}, G.~P. 2010, \apj, 708, 1760, \dodoi{10.1088/0004-637X/708/2/1760}

\bibitem[{{Getman} {et~al.}(2008{\natexlab{b}}){Getman}, {Feigelson}, {Micela},
  {Jardine}, {Gregory}, \& {Garmire}}]{Getman08b}
{Getman}, K.~V., {Feigelson}, E.~D., {Micela}, G., {et~al.} 2008{\natexlab{b}},
  \apj, 688, 437, \dodoi{10.1086/592034}

\bibitem[{{Getman} {et~al.}(2018){Getman}, {Kuhn}, {Feigelson}, {Broos},
  {Bate}, \& {Garmire}}]{Getman18b}
{Getman}, K.~V., {Kuhn}, M.~A., {Feigelson}, E.~D., {et~al.} 2018, \mnras, 477,
  298, \dodoi{10.1093/mnras/sty473}

\bibitem[{{Getman} {et~al.}(2005){Getman}, {Flaccomio}, {Broos}, {Grosso},
  {Tsujimoto}, {Townsley}, {Garmire}, {Kastner}, {Li}, {Harnden}, {Wolk},
  {Murray}, {Lada}, {Muench}, {McCaughrean}, {Meeus}, {Damiani}, {Micela},
  {Sciortino}, {Bally}, {Hillenbrand }, {Herbst}, {Preibisch}, \&
  {Feigelson}}]{Getman05}
{Getman}, K.~V., {Flaccomio}, E., {Broos}, P.~S., {et~al.} 2005, \apjs, 160,
  319, \dodoi{10.1086/432092}

\bibitem[{{Giardino} {et~al.}(2007{\natexlab{a}}){Giardino}, {Favata},
  {Micela}, {Sciortino}, \& {Winston}}]{Giardino2007}
{Giardino}, G., {Favata}, F., {Micela}, G., {Sciortino}, S., \& {Winston}, E.
  2007{\natexlab{a}}, \aap, 463, 275, \dodoi{10.1051/0004-6361:20066424}

\bibitem[{{Giardino} {et~al.}(2007{\natexlab{b}}){Giardino}, {Favata},
  {Pillitteri}, {Flaccomio}, {Micela}, \& {Sciortino}}]{Giardino2007b}
{Giardino}, G., {Favata}, F., {Pillitteri}, I., {et~al.} 2007{\natexlab{b}},
  \aap, 475, 891, \dodoi{10.1051/0004-6361:20077899}

\bibitem[{{Giardino} {et~al.}(2006){Giardino}, {Favata}, {Silva}, {Micela},
  {Reale}, \& {Sciortino}}]{Giardino2006}
{Giardino}, G., {Favata}, F., {Silva}, B., {et~al.} 2006, \aap, 453, 241,
  \dodoi{10.1051/0004-6361:20053663}

\bibitem[{{Gregory} {et~al.}(2016){Gregory}, {Adams}, \& {Davies}}]{Gregory16}
{Gregory}, S.~G., {Adams}, F.~C., \& {Davies}, C.~L. 2016, \mnras, 457, 3836,
  \dodoi{10.1093/mnras/stw259}

\bibitem[{{Gregory} {et~al.}(2014){Gregory}, {Donati}, {Morin}, {Hussain},
  {Mayne}, {Hillenbrand}, \& {Jardine}}]{Gregory2014}
{Gregory}, S.~G., {Donati}, J.~F., {Morin}, J., {et~al.} 2014, in Magnetic
  Fields throughout Stellar Evolution, ed. P.~{Petit}, M.~{Jardine}, \& H.~C.
  {Spruit}, Vol. 302, 40--43, \dodoi{10.1017/S1743921314001677}

\bibitem[{{Gregory} {et~al.}(2010){Gregory}, {Jardine}, {Gray}, \&
  {Donati}}]{Gregory2010}
{Gregory}, S.~G., {Jardine}, M., {Gray}, C.~G., \& {Donati}, J.~F. 2010,
  Reports on Progress in Physics, 73, 126901,
  \dodoi{10.1088/0034-4885/73/12/126901}

\bibitem[{{Gregory} {et~al.}(2007){Gregory}, {Wood}, \&
  {Jardine}}]{Gregory2007}
{Gregory}, S.~G., {Wood}, K., \& {Jardine}, M. 2007, \mnras, 379, L35,
  \dodoi{10.1111/j.1745-3933.2007.00328.x}

\bibitem[{{Grosso} {et~al.}(2020){Grosso}, {Hamaguchi}, {Principe}, \&
  {Kastner}}]{Grosso20}
{Grosso}, N., {Hamaguchi}, K., {Principe}, D.~A., \& {Kastner}, J.~H. 2020,
  \aap, 638, L4, \dodoi{10.1051/0004-6361/202038185}

\bibitem[{{Grosso} {et~al.}(1997){Grosso}, {Montmerle}, {Feigelson},
  {Andr{\'e}}, {Casanova}, \& {Gregorio-Hetem}}]{Grosso1997}
{Grosso}, N., {Montmerle}, T., {Feigelson}, E.~D., {et~al.} 1997, \nat, 387,
  56, \dodoi{10.1038/387056a0}

\bibitem[{{Grosso} {et~al.}(2004){Grosso}, {Montmerle}, {Feigelson}, \&
  {Forbes}}]{Grosso2004}
{Grosso}, N., {Montmerle}, T., {Feigelson}, E.~D., \& {Forbes}, T.~G. 2004,
  \aap, 419, 653, \dodoi{10.1051/0004-6361:20034070}

\bibitem[{{Guarcello} {et~al.}(2019){Guarcello}, {Micela}, {Sciortino},
  {L{\'o}pez-Santiago}, {Argiroffi}, {Reale}, {Flaccomio},
  {Alvarado-G{\'o}mez}, {Antoniou}, {Drake}, {Pillitteri}, {Rebull}, \&
  {Stauffer}}]{Guarcello2019}
{Guarcello}, M.~G., {Micela}, G., {Sciortino}, S., {et~al.} 2019, \aap, 622,
  A210, \dodoi{10.1051/0004-6361/201834370}

\bibitem[{{G{\"u}del}(2004)}]{Gudel2004}
{G{\"u}del}, M. 2004, \aapr, 12, 71, \dodoi{10.1007/s00159-004-0023-2}

\bibitem[{{G{\"u}del} {et~al.}(2007){G{\"u}del}, {Briggs}, {Arzner}, {Audard},
  {Bouvier}, {Feigelson}, {Franciosini}, {Glauser}, {Grosso}, {Micela},
  {Monin}, {Montmerle}, {Padgett}, {Palla}, {Pillitteri}, {Rebull}, {Scelsi},
  {Silva}, {Skinner}, {Stelzer}, \& {Telleschi}}]{Gudel2007b}
{G{\"u}del}, M., {Briggs}, K.~R., {Arzner}, K., {et~al.} 2007, \aap, 468, 353,
  \dodoi{10.1051/0004-6361:20065724}

\bibitem[{{Guedel} {et~al.}(1996){Guedel}, {Benz}, {Schmitt}, \&
  {Skinner}}]{Guedel1996}
{Guedel}, M., {Benz}, A.~O., {Schmitt}, J. H.~M.~M., \& {Skinner}, S.~L. 1996,
  \apj, 471, 1002, \dodoi{10.1086/178027}

\bibitem[{{Gully-Santiago} {et~al.}(2017){Gully-Santiago}, {Herczeg},
  {Czekala}, {Somers}, {Grankin}, {Covey}, {Donati}, {Alencar}, {Hussain},
  {Shappee}, {Mace}, {Lee}, {Holoien}, {Jose}, \& {Liu}}]{GullySantiago2017}
{Gully-Santiago}, M.~A., {Herczeg}, G.~J., {Czekala}, I., {et~al.} 2017, \apj,
  836, 200, \dodoi{10.3847/1538-4357/836/2/200}

\bibitem[{{Handy} {et~al.}(1999){Handy}, {Acton}, {Kankelborg}, {Wolfson},
  {Akin}, {Bruner}, {Caravalho}, {Catura}, {Chevalier}, {Duncan}, {Edwards},
  {Feinstein}, {Freeland }, {Friedlaender}, {Hoffmann}, {Hurlburt},
  {Jurcevich}, {Katz}, {Kelly}, {Lemen}, {Levay}, {Lindgren}, {Mathur},
  {Meyer}, {Morrison}, {Morrison}, {Nightingale}, {Pope}, {Rehse}, {Schrijver},
  {Shine}, {Shing}, {Strong}, {Tarbell}, {Title}, {Torgerson}, {Golub},
  {Bookbinder}, {Caldwell}, {Cheimets}, {Davis}, {Deluca}, {McMullen},
  {Warren}, {Amato}, {Fisher}, {Maldonado}, \& {Parkinson}}]{Handy1999}
{Handy}, B.~N., {Acton}, L.~W., {Kankelborg}, C.~C., {et~al.} 1999, \solphys,
  187, 229, \dodoi{10.1023/A:1005166902804}

\bibitem[{{Hayashi} {et~al.}(1996){Hayashi}, {Shibata}, \&
  {Matsumoto}}]{Hayashi1996}
{Hayashi}, M.~R., {Shibata}, K., \& {Matsumoto}, R. 1996, \apjl, 468, L37,
  \dodoi{10.1086/310222}

\bibitem[{{Hern{\'a}ndez} {et~al.}(2007){Hern{\'a}ndez}, {Hartmann}, {Megeath},
  {Gutermuth}, {Muzerolle}, {Calvet}, {Vivas}, {Brice{\~n}o}, {Allen},
  {Stauffer}, {Young}, \& {Fazio}}]{Hernandez2007}
{Hern{\'a}ndez}, J., {Hartmann}, L., {Megeath}, T., {et~al.} 2007, \apj, 662,
  1067, \dodoi{10.1086/513735}

\bibitem[{{Hubrig} {et~al.}(2015){Hubrig}, {Carroll}, {Scholler}, \&
  {Ilyin}}]{Hubrig2015}
{Hubrig}, S., {Carroll}, T.~A., {Scholler}, M., \& {Ilyin}, I. 2015, \mnras,
  449, L118, \dodoi{10.1093/mnrasl/slv034}

\bibitem[{{Jardine} {et~al.}(2006){Jardine}, {Collier Cameron}, {Donati},
  {Gregory}, \& {Wood}}]{Jardine2006}
{Jardine}, M., {Collier Cameron}, A., {Donati}, J.~F., {Gregory}, S.~G., \&
  {Wood}, K. 2006, \mnras, 367, 917, \dodoi{10.1111/j.1365-2966.2005.09995.x}

\bibitem[{{Johnstone} {et~al.}(2014){Johnstone}, {Jardine}, {Gregory},
  {Donati}, \& {Hussain}}]{Johnstone2014}
{Johnstone}, C.~P., {Jardine}, M., {Gregory}, S.~G., {Donati}, J.~F., \&
  {Hussain}, G. 2014, \mnras, 437, 3202, \dodoi{10.1093/mnras/stt2107}

\bibitem[{{Kirichenko} \& {Bogachev}(2017)}]{Kirichenko2017}
{Kirichenko}, A.~S., \& {Bogachev}, S.~A. 2017, \solphys, 292, 120,
  \dodoi{10.1007/s11207-017-1146-8}

\bibitem[{{Kochukhov} {et~al.}(2020){Kochukhov}, {Hackman}, {Lehtinen}, \&
  {Wehrhahn}}]{Kochukhov2020}
{Kochukhov}, O., {Hackman}, T., {Lehtinen}, J.~J., \& {Wehrhahn}, A. 2020,
  \aap, 635, A142, \dodoi{10.1051/0004-6361/201937185}

\bibitem[{{Koyama} {et~al.}(1996){Koyama}, {Hamaguchi}, {Ueno}, {Kobayashi}, \&
  {Feigelson}}]{Koyama1996}
{Koyama}, K., {Hamaguchi}, K., {Ueno}, S., {Kobayashi}, N., \& {Feigelson},
  E.~D. 1996, \pasj, 48, L87, \dodoi{10.1093/pasj/48.5.L87}

\bibitem[{{Lada} {et~al.}(2006){Lada}, {Muench}, {Luhman}, {Allen}, {Hartmann},
  {Megeath}, {Myers}, {Fazio}, {Wood}, {Muzerolle}, {Rieke}, {Siegler}, \&
  {Young}}]{Lada2006}
{Lada}, C.~J., {Muench}, A.~A., {Luhman}, K.~L., {et~al.} 2006, \aj, 131, 1574,
  \dodoi{10.1086/499808}

\bibitem[{Loader(1999)}]{Loader99}
Loader, C. 1999, Local Regression and Likelihood (Springer-Verlag New York),
  \dodoi{10.1007/b98858}

\bibitem[{{Loader}(2020)}]{Loader20}
{Loader}, C. 2020, locfit: Local Regression, Likelihood and Density Estimation,
   codes.
\newblock \url{https://cran.r-project.org/web/packages/locfit/index.html}

\bibitem[{{L{\'o}pez-Santiago} {et~al.}(2016){L{\'o}pez-Santiago},
  {Crespo-Chac{\'o}n}, {Flaccomio}, {Sciortino}, {Micela}, \&
  {Reale}}]{Lopez-Santiago2016}
{L{\'o}pez-Santiago}, J., {Crespo-Chac{\'o}n}, I., {Flaccomio}, E., {et~al.}
  2016, \aap, 590, A7, \dodoi{10.1051/0004-6361/201527499}

\bibitem[{{L{\'o}pez-Santiago} {et~al.}(2010){L{\'o}pez-Santiago},
  {Crespo-Chac{\'o}n}, {Micela}, \& {Reale}}]{LopezSantiago2010}
{L{\'o}pez-Santiago}, J., {Crespo-Chac{\'o}n}, I., {Micela}, G., \& {Reale}, F.
  2010, \apj, 712, 78, \dodoi{10.1088/0004-637X/712/1/78}

\bibitem[{{Massi} {et~al.}(2008){Massi}, {Ros}, {Menten}, {Kaufman
  Bernad{\'o}}, {Torricelli-Ciamponi}, {Neidh{\"o}fer}, {Boden}, {Boboltz},
  {Sargent}, \& {Torres}}]{Massi2008}
{Massi}, M., {Ros}, E., {Menten}, K.~M., {et~al.} 2008, \aap, 480, 489,
  \dodoi{10.1051/0004-6361:20078637}

\bibitem[{{McCleary} \& {Wolk}(2011)}]{McCleary2011}
{McCleary}, J.~E., \& {Wolk}, S.~J. 2011, \aj, 141, 201,
  \dodoi{10.1088/0004-6256/141/6/201}

\bibitem[{{Megeath} {et~al.}(2012){Megeath}, {Gutermuth}, {Muzerolle},
  {Kryukova}, {Flaherty}, {Hora}, {Allen}, {Hartmann}, {Myers}, {Pipher},
  {Stauffer}, {Young}, \& {Fazio}}]{Megeath2012}
{Megeath}, S.~T., {Gutermuth}, R., {Muzerolle}, J., {et~al.} 2012, \aj, 144,
  192, \dodoi{10.1088/0004-6256/144/6/192}

\bibitem[{{Mewe} {et~al.}(1985){Mewe}, {Gronenschild}, \& {van den
  Oord}}]{Mewe1985}
{Mewe}, R., {Gronenschild}, E.~H.~B.~M., \& {van den Oord}, G.~H.~J. 1985,
  \aaps, 62, 197

\bibitem[{{Morris}(2020)}]{Morris2020}
{Morris}, B.~M. 2020, \apj, 893, 67, \dodoi{10.3847/1538-4357/ab79a0}

\bibitem[{{Moschou} {et~al.}(2019){Moschou}, {Drake}, {Cohen},
  {Alvarado-G{\'o}mez}, {Garraffo}, \& {Fraschetti}}]{Moschou2019}
{Moschou}, S.-P., {Drake}, J.~J., {Cohen}, O., {et~al.} 2019, \apj, 877, 105,
  \dodoi{10.3847/1538-4357/ab1b37}

\bibitem[{{Okamoto} {et~al.}(2021){Okamoto}, {Notsu}, {Maehara}, {Namekata},
  {Honda}, {Ikuta}, {Nogami}, \& {Shibata}}]{Okamoto2021}
{Okamoto}, S., {Notsu}, Y., {Maehara}, H., {et~al.} 2021, \apj, 906, 72,
  \dodoi{10.3847/1538-4357/abc8f5}

\bibitem[{{Pevtsov} {et~al.}(2003){Pevtsov}, {Fisher}, {Acton}, {Longcope},
  {Johns-Krull}, {Kankelborg}, \& {Metcalf}}]{Pevtsov2003}
{Pevtsov}, A.~A., {Fisher}, G.~H., {Acton}, L.~W., {et~al.} 2003, \apj, 598,
  1387, \dodoi{10.1086/378944}

\bibitem[{{Pillitteri} {et~al.}(2019){Pillitteri}, {Sciortino}, {Reale},
  {Micela}, {Argiroffi}, {Flaccomio}, \& {Stelzer}}]{Pillitteri2019}
{Pillitteri}, I., {Sciortino}, S., {Reale}, F., {et~al.} 2019, \aap, 623, A67,
  \dodoi{10.1051/0004-6361/201834204}

\bibitem[{{Preibisch} {et~al.}(2005){Preibisch}, {Kim}, {Favata}, {Feigelson},
  {Flaccomio}, {Getman}, {Micela}, {Sciortino}, {Stassun}, {Stelzer}, \&
  {Zinnecker}}]{Preibisch05}
{Preibisch}, T., {Kim}, Y.-C., {Favata}, F., {et~al.} 2005, \apjs, 160, 401,
  \dodoi{10.1086/432891}

\bibitem[{{R Core Team}(2020)}]{RCoreTeam20}
{R Core Team}. 2020, R: A language and environment for statistical computing,
  R Foundation for Statistical Computing, Vienna, Austria.
\newblock \url{https://www.R-project.org}

\bibitem[{{Reale}(2007)}]{Reale2007}
{Reale}, F. 2007, \aap, 471, 271, \dodoi{10.1051/0004-6361:20077223}

\bibitem[{{Reale}(2014)}]{Reale2014}
---. 2014, Living Reviews in Solar Physics, 11, 4, \dodoi{10.12942/lrsp-2014-4}

\bibitem[{{Reale} {et~al.}(1997){Reale}, {Betta}, {Peres}, {Serio}, \&
  {McTiernan}}]{Reale1997}
{Reale}, F., {Betta}, R., {Peres}, G., {Serio}, S., \& {McTiernan}, J. 1997,
  \aap, 325, 782

\bibitem[{{Reale} {et~al.}(2004){Reale}, {G{\"u}del}, {Peres}, \&
  {Audard}}]{Reale2004}
{Reale}, F., {G{\"u}del}, M., {Peres}, G., \& {Audard}, M. 2004, \aap, 416,
  733, \dodoi{10.1051/0004-6361:20034027}

\bibitem[{{Reale} {et~al.}(2018){Reale}, {Lopez-Santiago}, {Flaccomio},
  {Petralia}, \& {Sciortino}}]{Reale2018}
{Reale}, F., {Lopez-Santiago}, J., {Flaccomio}, E., {Petralia}, A., \&
  {Sciortino}, S. 2018, \apj, 856, 51, \dodoi{10.3847/1538-4357/aaaf1f}

\bibitem[{{Romanova} \& {Owocki}(2015)}]{Romanova2015}
{Romanova}, M.~M., \& {Owocki}, S.~P. 2015, \ssr, 191, 339,
  \dodoi{10.1007/s11214-015-0200-9}

\bibitem[{{Rosner} {et~al.}(1978){Rosner}, {Tucker}, \& {Vaiana}}]{Rosner1978}
{Rosner}, R., {Tucker}, W.~H., \& {Vaiana}, G.~S. 1978, \apj, 220, 643,
  \dodoi{10.1086/155949}

\bibitem[{{Scargle} {et~al.}(2013){Scargle}, {Norris}, {Jackson}, \&
  {Chiang}}]{Scargle2013}
{Scargle}, J.~D., {Norris}, J.~P., {Jackson}, B., \& {Chiang}, J. 2013, \apj,
  764, 167, \dodoi{10.1088/0004-637X/764/2/167}

\bibitem[{{Schmitt} {et~al.}(1985){Schmitt}, {Golub}, {Harnden}, {Maxson},
  {Rosner}, \& {Vaiana}}]{Schmitt1985}
{Schmitt}, J.~H.~M.~M., {Golub}, L., {Harnden}, F.~R., J., {et~al.} 1985, \apj,
  290, 307, \dodoi{10.1086/162986}

\bibitem[{{Serio} {et~al.}(1981){Serio}, {Peres}, {Vaiana}, {Golub}, \&
  {Rosner}}]{Serio1981}
{Serio}, S., {Peres}, G., {Vaiana}, G.~S., {Golub}, L., \& {Rosner}, R. 1981,
  \apj, 243, 288, \dodoi{10.1086/158597}

\bibitem[{{Serio} {et~al.}(1991){Serio}, {Reale}, {Jakimiec}, {Sylwester}, \&
  {Sylwester}}]{Serio1991}
{Serio}, S., {Reale}, F., {Jakimiec}, J., {Sylwester}, B., \& {Sylwester}, J.
  1991, \aap, 241, 197

\bibitem[{{Shibata} \& {Magara}(2011)}]{Shibata2011}
{Shibata}, K., \& {Magara}, T. 2011, Living Reviews in Solar Physics, 8, 6,
  \dodoi{10.12942/lrsp-2011-6}

\bibitem[{{Shu} {et~al.}(1997){Shu}, {Shang}, {Glassgold}, \& {Lee}}]{Shu1997}
{Shu}, F.~H., {Shang}, H., {Glassgold}, A.~E., \& {Lee}, T. 1997, Science, 277,
  1475, \dodoi{10.1126/science.277.5331.1475}

\bibitem[{{Sokal} {et~al.}(2020){Sokal}, {Johns-Krull}, {Mace}, {Nofi},
  {Prato}, {Lee}, \& {Jaffe}}]{Sokal2020}
{Sokal}, K.~R., {Johns-Krull}, C.~M., {Mace}, G.~N., {et~al.} 2020, \apj, 888,
  116, \dodoi{10.3847/1538-4357/ab59d8}

\bibitem[{{Stelzer} {et~al.}(2007){Stelzer}, {Flaccomio}, {Briggs}, {Micela},
  {Scelsi}, {Audard}, {Pillitteri}, \& {G{\"u}del}}]{Stelzer07}
{Stelzer}, B., {Flaccomio}, E., {Briggs}, K., {et~al.} 2007, \aap, 468, 463,
  \dodoi{10.1051/0004-6361:20066043}

\bibitem[{{Stelzer} {et~al.}(2005){Stelzer}, {Flaccomio}, {Montmerle},
  {Micela}, {Sciortino}, {Favata}, {Preibisch}, \& {Feigelson}}]{Stelzer05}
{Stelzer}, B., {Flaccomio}, E., {Montmerle}, T., {et~al.} 2005, \apjs, 160,
  557, \dodoi{10.1086/432375}

\bibitem[{{Stelzer} {et~al.}(2009){Stelzer}, {Robrade}, {Schmitt}, \&
  {Bouvier}}]{Stelzer09}
{Stelzer}, B., {Robrade}, J., {Schmitt}, J.~H.~M.~M., \& {Bouvier}, J. 2009,
  \aap, 493, 1109, \dodoi{10.1051/0004-6361:200810540}

\bibitem[{{Su} {et~al.}(2007){Su}, {Van Ballegooijen}, {McCaughey}, {Deluca},
  {Reeves}, \& {Golub}}]{Su2007}
{Su}, Y., {Van Ballegooijen}, A., {McCaughey}, J., {et~al.} 2007, \apj, 665,
  1448, \dodoi{10.1086/519679}

\bibitem[{{Takasao} {et~al.}(2019){Takasao}, {Tomida}, {Iwasaki}, \&
  {Suzuki}}]{Takasao2019}
{Takasao}, S., {Tomida}, K., {Iwasaki}, K., \& {Suzuki}, T.~K. 2019, \apjl,
  878, L10, \dodoi{10.3847/2041-8213/ab22bb}

\bibitem[{{Telleschi} {et~al.}(2007){Telleschi}, {G{\"u}del}, {Briggs},
  {Audard}, \& {Palla}}]{Telleschi07}
{Telleschi}, A., {G{\"u}del}, M., {Briggs}, K.~R., {Audard}, M., \& {Palla}, F.
  2007, \aap, 468, 425, \dodoi{10.1051/0004-6361:20066565}

\bibitem[{{Unruh} \& {Jardine}(1997)}]{Unruh1997}
{Unruh}, Y.~C., \& {Jardine}, M. 1997, \aap, 321, 177

\bibitem[{{Vidotto} {et~al.}(2009){Vidotto}, {Opher}, {Jatenco-Pereira}, \&
  {Gombosi}}]{Vidotto2009}
{Vidotto}, A.~A., {Opher}, M., {Jatenco-Pereira}, V., \& {Gombosi}, T.~I. 2009,
  \apj, 703, 1734, \dodoi{10.1088/0004-637X/703/2/1734}

\bibitem[{{Vidotto} {et~al.}(2014){Vidotto}, {Gregory}, {Jardine}, {Donati},
  {Petit}, {Morin}, {Folsom}, {Bouvier}, {Cameron}, {Hussain}, {Marsden},
  {Waite}, {Fares}, {Jeffers}, \& {do Nascimento}}]{Vidotto2014}
{Vidotto}, A.~A., {Gregory}, S.~G., {Jardine}, M., {et~al.} 2014, \mnras, 441,
  2361, \dodoi{10.1093/mnras/stu728}

\bibitem[{{Warnecke} \& {K{\"a}pyl{\"a}}(2020)}]{Warnecke2020}
{Warnecke}, J., \& {K{\"a}pyl{\"a}}, M.~J. 2020, \aap, 642, A66,
  \dodoi{10.1051/0004-6361/201936922}

\bibitem[{{Wolk} {et~al.}(2005){Wolk}, {Harnden}, {Flaccomio}, {Micela},
  {Favata}, {Shang}, \& {Feigelson}}]{Wolk05}
{Wolk}, S.~J., {Harnden}, F.~R., J., {Flaccomio}, E., {et~al.} 2005, \apjs,
  160, 423, \dodoi{10.1086/432099}

\bibitem[{{Yadav} {et~al.}(2015){Yadav}, {Christensen}, {Morin}, {Gastine},
  {Reiners}, {Poppenhaeger}, \& {Wolk}}]{Yadav2015}
{Yadav}, R.~K., {Christensen}, U.~R., {Morin}, J., {et~al.} 2015, \apjl, 813,
  L31, \dodoi{10.1088/2041-8205/813/2/L31}

\bibitem[{{Yang} \& {Johns-Krull}(2011)}]{Yang2011}
{Yang}, H., \& {Johns-Krull}, C.~M. 2011, \apj, 729, 83,
  \dodoi{10.1088/0004-637X/729/2/83}

\bibitem[{{Zhuleku} {et~al.}(2020){Zhuleku}, {Warnecke}, \&
  {Peter}}]{Zhuleku2020}
{Zhuleku}, J., {Warnecke}, J., \& {Peter}, H. 2020, \aap, 640, A119,
  \dodoi{10.1051/0004-6361/202038022}

\bibitem[{{Zhuleku} {et~al.}(2021){Zhuleku}, {Warnecke}, \&
  {Peter}}]{Zhuleku2021}
---. 2021, arXiv e-prints, arXiv:2102.00982.
\newblock \doarXiv{2102.00982}

\end{thebibliography}
\bibliographystyle{aasjournal}
\end{document}